\documentclass[seceq]{ptptex}
\usepackage{amsmath}

\begin{document}
\pagestyle{plain}


\begin{flushright}
RUP-06-1 \\ 
hep-th/0608185
\end{flushright}

\baselineskip=7mm
\renewcommand{\baselinestretch}{1.28}
\newcommand{\ba}{\begin{eqnarray}}
\newcommand{\ea}{\end{eqnarray}}
\newcommand{\non}{\nonumber\\}

\hskip 12.36cm{}
\vspace{3cm}

\centerline{\LARGE  Covariant BRST Quantization of Closed Strings}
\vskip 2mm
\centerline{\LARGE in the {{\it PP}-Wave Background}}

\vspace{1.5cm}

\centerline{\bf Yoichi Chizaki
\footnote{E-mail: {\tt yoichizaki@stu.rikkyo.ne.jp}}
and Shigeaki Yahikozawa
\footnote{
E-mail: {\tt yahiko@rikkyo.ac.jp}}}
\vspace{1mm}

\centerline{\it Department of Physics, Rikkyo University, Tokyo 171-8501, Japan}

\vspace{1cm}

\centerline{\bf Abstract}
\vspace{2mm}
We canonically quantize closed string theory in the {\it pp}-wave
background with a non-zero flux of the three-form field strength
by using the covariant BRST operator formalism. 
In this canonical quantization, 
we completely construct new covariant free-mode representations, 
for which it is particularly important to take account of the commutation relations 
of the zero mode of the light-cone string coordinate 
$X^{-}$ with other modes. 
All covariant string coordinates are composed of free modes. 
Moreover, employing these covariant string coordinates for 
the energy-momentum tensor, we calculate the anomaly in the Virasoro algebra 
and determine the number of dimensions of spacetime and the ordering constant 
from the nilpotency condition of the BRST charge 
in the {\it pp}-wave background.


\newpage
\section{Introduction}

Understanding the quantization of strings in a variety of
backgrounds is of great importance. 
In particular, this quantization is significant for analyzing
the string landscape, the AdS/CFT correspondence, matrix models 
and string phenomenology.
Of course, quantization in a number of interesting backgrounds has already 
been investigated and has been applied to many models.
One of the interesting backgrounds with respect to which quantization
has been studied is the {\it pp}-wave background.
Bosonic string theory in the {\it pp}-wave background has been
quantized through application of the canonical operator formalism in the 
light-cone gauge,
\cite{Horowitz-Steif-1}\tocite{Tseytlin95}
and it has been investigated
by the way of the path integral formalism\cite{Amati-Klim,Rudd}
and from the point of view of 
the world-sheet conformal field theory \cite{Kiritsis93}.
The quantization of superstrings in the {\it pp}-wave background with 
a non-zero flux of the RR five-form field strength has been carried out
in the light-cone gauge,\cite{Metsaev,M-T}
and this quantization has been used in the BMN correspondence\cite{BMN}.
Moreover, in the covariant quantization of
superstrings in the NS-NS {\it pp}-wave background, a free field realization 
of current algebra and  the Sugawara construction of 
the world-sheet conformal field theory 
have been used. \cite{Kiritsis94,Kunitomo}
Although the methods mentioned above are very useful, 
there is another method important for understanding the structure of spacetime
and constructing a covariant string field theory in the {\it pp}-wave 
background.
That is, it is important to consider the covariant BRST quantization
in the {\it pp}-wave background from the point of view of 
the canonical operator formalism.
From this covariant formalism, we should be able to understand
the covariant BMN correspondence.
In addition, it is necessary to elucidate how all approaches are related.

In this paper, we canonically quantize a closed bosonic string in 
the {\it pp}-wave background with a non-zero flux of the three-form 
field strength of the antisymmetric two-form field by using the 
covariant BRST operator formalism.
First, we construct new free-mode representations of
all the covariant string coordinates.
Here, we would like to emphasize that these covariant string coordinates
in the free-mode representation must satisfy the condition of the canonical 
commutation relations and must be general solutions of 
the Heisenberg equations of motion whose form is that of
the Euler-Lagrange equations of motion in the {\it pp}-wave background.
Second, by using the free-mode representations of
the covariant string coordinates for the energy-momentum tensor, 
we calculate the anomaly in the Virasoro algebra 
and determine the number of dimensions of spacetime and the ordering constant 
from the nilpotency condition of the BRST charge 
in the {\it pp}-wave background.

This paper is organized as follows.
In \S2 we briefly review the action and the general solutions of the
equations of motion of a closed string in the {\it pp}-wave background
and define our notation.
Moreover, we review the quantization of ghosts and antighosts.
In \S3 we present new free-mode representations of 
all the covariant string coordinates.
The free-mode representation of a light-cone string coordinate 
$X^{-}$  is characteristic.
In \S4 we prove that the free-mode representations satisfy
all the equal-time canonical commutation relations among
all the covariant string coordinates.
In \S5 we calculate the anomaly in the Virasoro algebra by using
the energy-momentum tensor in the free-mode representation of
all the covariant string coordinates.
In \S6 we determine the number of dimensions of spacetime and the ordering 
constant from the nilpotency condition of the BRST charge 
in the {\it pp}-wave background.
Section 7 contains some conclusions.
In Appendix A we construct new  general classical solutions of
a closed string in the {\it pp}-wave background
without the antisymmetric tensor field, 
while Appendix B contains some details of the special
mode expansion.


\vskip 1cm
\section{Notation and review}

We begin by defining our notation and reviewing the action, the background,
the equations of motion, the general solutions and the quantization of ghosts.
In \S2.1 we define the total action of the closed bosonic string in the
{\it pp}-wave background with the flux of the antisymmetric tensor field
in $D$ spacetime dimensions; here we ignore the dilaton field. 
In \S2.2 we explain the equations of motion of strings and their 
general solutions in the {\it pp}-wave background with a flux. 
In \S2.3 we review the equations of motion of ghosts, their
general solutions and the quantization of ghosts.


\vskip 1cm
\subsection{Action and backgrounds}

Our starting point is the total action $S=S_{X}+S_{\rm{GF+gh}}$,
which is BRST invariant:
\begin{align}
S_{X}&=\frac{-1}{4\pi\alpha'}\int d\tau d\sigma
\left[
\sqrt{-g}g^{ab}G_{\mu\nu}
+\epsilon^{ab}B_{\mu\nu}
\right]
\partial_{a}X^{\mu}\partial_{b}X^{\nu}\,\,,
\label{Action-X-first}\\
S_{\rm{GF+gh}}&=\int d\tau d\sigma\sqrt{-g}
\left[
\frac{1}{4\pi}{\cal B}_{ab}\left(g^{ab}-\eta^{ab}\right)
-\frac{i}{2\pi}b_{ab}\nabla^{a}c^{b}
\right]\,\,,
\end{align}
where $g_{ab}$, $\epsilon^{ab}$, and  $\eta^{ab}$ are, respectively,  a general 
world-sheet metric, the totally world-sheet antisymmetric tensor 
($\epsilon^{01}=+1$), and the flat world-sheet metric, which is diag$(-1, +1)$. 
In the nonlinear sigma model action $S_{X}$, $G_{\mu\nu}$ and $B_{\mu\nu}$
are, respectively, a general spacetime string metric and an antisymmetric tensor field,
and the spacetime indices $\mu$ and $\nu$ run over 
$+\,,\,-\,,\,2\,,\,3\,,\,\cdots\,,\,D-1$.
In the action $S_{\rm{GF+gh}} =S_{\rm{GF}}+S_{\rm{gh}}$,
where $S_{\rm{GF}}$ is the gauge fixing action and
$S_{\rm{gh}}$ is the Faddeev-Popov ghost action, ${\cal B}_{ab}$, $c^{a}$, 
and $b_{ab}$ are, respectively, the auxiliary field to fix  the gauge, the ghost field, 
and the antighost field.
Because we construct the covariant BRST quantization for string theory in this paper, 
we choose the covariant gauge-fixing condition on the world-sheet,
$g^{ab}=\eta^{ab}$. This covariant gauge-fixing condition is obtained from 
the equation of motion for the auxiliary field, ${\cal B}_{ab}$.
After the gauge fixing, we use the world-sheet light-cone coordinates 
$\sigma^{\pm}=\tau\pm\sigma$, so that 
the components of the world-sheet metric and 
the world-sheet totally antisymmetric tensor become
$\eta^{+-}=\eta^{-+}=-2$ and $\epsilon^{+-}=-\epsilon^{-+}=-2$.
Also, their partial derivatives are then 
$\partial_{\pm}=\frac{1}{2}(\partial_{\tau}\pm\partial_{\sigma})$.
Moreover, we use the spacetime light-cone coordinates
$X^{\pm}=\frac{1}{\sqrt{2}}(X^{0}\pm X^{1})$.

The condition that string theory be Weyl-invariant in its quantization on
the world-sheet requires that the renormalization group $\beta$-functions
vanish at all loop orders; 
these necessary conditions correspond to the field equations,
which resemble Einstein's equation, the antisymmetric
tensor generalization of Maxwell's equation, and so on\cite{Polchinski}.
As the background field that which satisfies these field equations, 
we use the following $\it pp$-wave metric and antisymmetric
tensor field, whose flux is a constant:
\begin{align}
ds^2&=-\mu^2(X^2+Y^2)dX^{+}dX^{+}-2dX^{+}dX^{-}+ dXdX+dYdY
+dX^{k}dX^{k}\,\,, \\
B&=-\mu YdX^{+}\wedge dX+\mu XdX^{+}\wedge dY\,\,.
\end{align}
Here, we define $X^{\mu=2}=X$ and $X^{\mu=3}=Y$, 
and the index $k$ runs over $4\,,\,5\,,\,\cdots\,,\,D-1$. 
Thus, the components of $G_{\mu\nu}$ and $B_{\mu\nu}$ are
\begin{align}
G_{++}&=-\mu^2(X^2+Y^2)\,,
\,\,\,\,\,
G_{+-}=G_{-+}=-1 \,, \\
G_{ij}&=\delta_{ij}\,,\,\,\,\,\,\, i,j=2,3,\cdots, D-1\,, \\
B_{+2}&=-B_{2+}=-\mu\,Y
\,,\,\,\,\,\,\,\,\,
B_{+3}=-B_{3+}=\mu\,X\,,
\end{align}
with all others vanishing.
This is almost identical to the Nappi-Witten background \cite{Nappi-Witten}.

Finally, we introduce the complex coordinates $Z=X+iY$ and $\bar{Z}=X-iY$.
Then the action $S_{X}$ takes the simple form
\begin{align}
S_{X}=\frac{1}{2\pi\alpha'}\int d\tau d\sigma
\Bigl[
&-2\partial_{+}X^{+}\partial_{-}X^{-}-2\partial_{-}X^{+}\partial_{+}X^{-}
\nonumber\\
&+\bar{\cal D}_{+}\bar{Z}{\cal D}_{-}Z
+\bar{{\cal D}}_{-}\bar{Z}{\cal D}_{+}Z+2\partial_{+}X^{k}\partial_{-}X^{k}
\Bigr]\,,
\label{action-X}
\end{align}
where the world-sheet covariant derivatives are defined as
${\cal D}_{\pm}=\partial_{\pm}\pm i\mu\partial_{\pm}X^{+}$,
which have a form similar to the covariant derivatives of quantum electrodynamics.
After integrating out the auxiliary field ${\cal B}_{ab}$, 
the gauge fixing action $S_{\rm{GF}}$ vanishes, 
and the Faddeev-Popov ghost action $S_{\rm gh}$ is reduced to 
\begin{align}
S_{\rm gh}=\frac{i}{\pi}\int d\tau d\sigma 
\bigl[
b_{++}\partial_{-}c^{+}
+b_{--}\partial_{+}c^{-}
\bigr]\,.
\label{ghost-action}
\end{align}


\vskip 1cm
\subsection{The equations of motion of $X^{\mu}$ and their general solutions}

We obtain the equations of motion of $X^{\mu}$ from 
the action (\ref{action-X}). 
These equations are obviously related to the Heisenberg equations 
of motion with respect to quantization.

$\bullet\,\,$ The equations of motion of $X^{+}$ and $X^{k}$ are
\begin{align}
\partial_{+}\partial_{-}X^{+}=0\,,
\,\,\,\,\,
\quad\partial_{+}\partial_{-}X^{k}=0\,.
\label{eom:X+k}
\end{align}

$\bullet\,\,$ The equations of motion of $Z$ and $\bar{Z}$ are
\begin{align}
{\cal D}_{+}{\cal D}_{-}Z=0\,,
\,\,\,\,\,
\quad\bar{\cal D}_{+}\bar{\cal D}_{-}\bar{Z}=0\,.
\label{eom:Z}
\end{align}

$\bullet\,\,$ The equation of motion of $X^{-}$ is
\begin{align}
\partial_{+}\partial_{-}X^{-}+\frac{i\mu}{4}
\left[
\partial_{+}\left(\bar{Z}{\cal D}_{-}Z
-Z\bar{\cal D}_{-}\bar{Z}\right)-
\partial_{-}\left(\bar{Z}{\cal D}_{+}Z-Z\bar{\cal D}_{+}\bar{Z}\right)
\right]
=0\,.
\label{eom:X-}
\end{align}
In Eq.(\ref{eom:X+k}), $X^{+}$ and $X^{k}$ satisfy the free field equations. 
By contrast, $Z$ and $\bar{Z}$ interact with 
$X^{+}$ through the covariant derivatives in Eq.(\ref{eom:Z}),
and $X^{-}$ interacts with $Z$, $\bar{Z}$ and $X^{+}$ in Eq.(\ref{eom:X-}).
Nevertheless, we are able to obtain general solutions for all $X^{\mu}$
in the following.

First, we simply solve the equations of motion for the free fields
$X^{+}$ and $X^{k}$ under the periodic condition of closed string theory.
In this way, we find that the general solutions to these equations are 
the normal d'Alembert solutions,
\begin{align}
X^{+}=X^{+}_{\rm L}(\sigma^{+})+X^{+}_{\rm R}(\sigma^{-})\,,
\quad \,\,\,
X^{k}=X^{k}_{\rm L}(\sigma^{+})+X^{k}_{\rm R}(\sigma^{-})\,,
\label{sol:X+}
\end{align}
where L  and R indicate the left-moving and right-moving parts, respectively.
Second, we solve the equation of motion for $Z$ under the periodic
condition of closed string theory. 
To accomplish this, we define $\tilde{X}^{+} \equiv X^{+}_{\rm L}-X^{+}_{\rm R}$ and 
multiply the equation of motion for $Z$, Eq.(\ref{eom:Z}),
by $e^{i\mu\tilde{X}^{+}}$ from the left.
Then, the equation of motion for $Z$ becomes
\begin{align}
\partial_{+}\partial_{-}\left[e^{i\mu\tilde{X}^{+}}Z\right]=0.
\label{eq-new-Z}
\end{align}
This equation shows that $e^{i\mu\tilde{X}^{+}}Z$ is similar to a free field.
Therefore, the quantity $e^{i\mu\tilde{X}^{+}}Z$ can be expressed as
a sum of arbitrary functions $f(\sigma^{+})$ and $g(\sigma^{-})$
that satisfy the twisted boundary conditions, 
and the general form of $Z$ is
\begin{align}
Z=e^{-i\mu\tilde{X}^{+}}
\bigl[
f(\sigma^{+})+g(\sigma^{-})
\bigr]\,.
\label{sol:Z}
\end{align}
Since the general form of $\bar{Z}$ can be obtained 
from the complex conjugate of $Z$, we have
\begin{align}
\bar{Z}=e^{i\mu\tilde{X}^{+}}
\bigl[
\bar{f}(\sigma^{+})+\bar{g}(\sigma^{-})
\bigr]\,.
\label{sol:Zbar}
\end{align}
Finally, in order to solve the equation of motion for $X^{-}$ under the periodic 
condition of closed string theory, we substitute the solutions $Z$ given in
Eq.(\ref{sol:Z}) and $\bar{Z}$ given in Eq.(\ref{sol:Zbar}) into
the equation of motion for $X^{-}$, Eq.(\ref{eom:X-}).
In this way, the equation of motion for $X^{-}$ is reduced to a simpler form,
\begin{align}
\partial_{+}\partial_{-}X^{-}
-\frac{i\mu}{2}
\left[
\partial_{+}f\partial_{-}\bar{g}
-\partial_{+}\bar{f}\partial_{-}g
\right]
=0\,.
\label{eq-new-X^{-}}
\end{align}
The important point here is that Eq.(\ref{eq-new-X^{-}}) does not include
$X^{+}$.
Moreover, because $f$ is an arbitrary function of $\sigma^{+}$ and 
$g$ is an arbitrary function of $\sigma^{-}$, 
we have $\partial_{+}f\partial_{-}\bar{g}-\partial_{+}\bar{f}\partial_{-}g
=\partial_{+}\partial_{-}(f\bar{g}-\bar{f}g)$.
In terms of this relation, Eq.(\ref{eq-new-X^{-}}) becomes
\begin{align}
\partial_{+}\partial_{-}
\left[
X^{-}-\frac{i\mu}{2}
\left(
f\bar{g}-\bar{f}g
\right)
\right]
=0\,.
\label{eq-fin-X^{-}}
\end{align}
Because this equation is of a classical free field type,
similar to the equation of motion for $Z$ Eq.(\ref{eq-new-Z}),
we can easily solve Eq.(\ref{eq-fin-X^{-}}).
Doing so, we obtain its general solution,
\begin{align}
X^{-}=X^{-}_{\rm L}(\sigma^{+})+X^{-}_{\rm R}(\sigma^{-})
+\frac{i\mu}{2}
\left(
f\bar{g}-\bar{f}g
\right)\,,
\label{sol:X-} 
\end{align}
where $X^{-}_{\rm L}$ is an arbitrary function of $\sigma^{+}$, 
$X^{-}_{\rm R}$ is an arbitrary function of $\sigma^{-}$,
and $X^{-}_{\rm L}+X^{-}_{\rm R}$ is 
a periodic function of $\sigma$.
Note that $X^{-}_{\rm L}$ and $X^{-}_{\rm R}$ are not free fields
in the quantized theory,
although $X^{+}_{\rm L}$, $X^{+}_{\rm R}$, $X^{k}_{\rm L}$ 
and $X^{k}_{\rm R}$ are completely free fields in the quantized theory.
Moreover, it is important that 
$X^{-}_{\rm L}$ and $X^{-}_{\rm R}$ can be divided into
almost free parts and completely non-free parts.
We give detailed discussion of these points in \S 3.

We explain closed string theory in the {\it pp}-wave background without
a non-zero flux of the antisymmetric tensor field 
(i.e., $B_{\mu\nu}=0$ ) in Appendix A.
There, constructing the action and the equations of motion in the covariant gauge, 
we find the general solutions and the energy momentum tensor. 
In particular, the general solutions $X^{\mu}$ given in that appendix 
are new solutions, and for this reason, the new mode expansion is significant.


\vskip 1cm
\subsection{The equations of motion and the quantization of the ghost system}

We obtain the equations of motion for ghosts and antighosts
from the action (\ref{ghost-action}):
\begin{align}
\partial_{-}c^{+}=0\,, \qquad  \partial_{+}c^{-}=0\,, \qquad
\partial_{-}b_{++}=0\,,\qquad  \partial_{+}b_{--}=0\,.
\end{align}
The general solutions $c^{+}$ and $b_{++}$
are purely left-moving, whereas the general solutions $c^{-}$ and $b_{--}$
are purely right-moving.
For closed string theory, the ghosts and antighosts satisfy 
a periodic condition, which is simply periodicity in $\sigma$ of period $2\pi$.
Therefore,  $c^{+}$ and $c^{-}$ have independent mode expansions.
Similarly, $b_{++}$ and $b_{--}$ also have independent mode expansions.
Thus the mode expansions of the ghosts and the antighosts are
\begin{align}
c^{+}&=\sum_{n}\tilde{c}_{n}e^{-in\sigma^{+}},
\,\,\,\,\,\,\,\,\,\,
\ c^{-}=\sum_{n}c_{n}e^{-in\sigma^{-}},
\label{solution:c}\\
b_{++}&=\sum_{n}\tilde{b}_{n}e^{-in\sigma^{+}},
\,\,\,\,\,\,\,
\ b_{--}=\sum_{n}b_{n}e^{-in\sigma^{-}}.
\label{solution:b}
\end{align}
The ghost system is quantized 
according to the following equal-time canonical anticommutation relations:
\begin{align}
\left\{c^{+}(\tau,\sigma),b_{++}(\tau,\sigma')\right\}=2\pi\delta(\sigma-\sigma')\,,
\qquad
\left\{c^{-}(\tau,\sigma),b_{--}(\tau,\sigma')\right\}=2\pi\delta(\sigma-\sigma')\,,
\end{align}
with all other anticommutators vanishing.
In terms of the modes, the anticommutation relations are
\begin{align}
\{\tilde{c}_{m},\tilde{b}_{n}\}=&\{c_{m},b_{n}\}=\delta_{m+n}\,, \\
\{\tilde{c}_{m},\tilde{c}_{n}\}=\{c_{m},c_{n}\}=0\,, &
\qquad \quad
\{\tilde{b}_{m},\tilde{b}_{n}\}=\{b_{m},b_{n}\}=0\,,
\end{align}
with the anticommutators of the left modes with the right
modes vanishing.\cite{G-S-W}


\vskip 1cm
\section{Free-mode representation}

In this section we derive the free-mode representations of all the
covariant string coordinates in closed string theory
in the {\it pp}-wave background with a non-zero flux of the three-form field 
strength of the antisymmetric tensor field.
In particular, we place special emphasis on the new free-mode representations 
of $Z$, $\bar{Z}$ and $X^{-}$.
These new free-mode representations satisfy 
the quantum condition that 
all the equal-time canonical commutation relations 
and the Heisenberg equations of motion must be satisfied.
The proof of the free-mode representations in the canonical quantization 
is given in the next section.
When constructing the free-mode representations, 
it is important to express the general solutions 
in terms of free-mode expansions and to consider 
the commutation relations of the zero mode of $X^{-}$ with other modes. 
Of course,  the commutators between the modes 
of different fields must completely vanish in the free-mode expansions.
The free-mode representations are useful for the calculation of 
the anomaly of the Virasoro algebra, the construction of the BRST
quantization, and so on.
Moreover, the free-mode representations may be effective for the purpose
of investigating the exact formation of the free-field representation.

We now present the free-mode representations of $X^{+}$, $X^{k}$,
$Z$, $\bar{Z}$ and $X^{-}$ in a clear manner. 
First, the free-mode representations of $X^{+}$ and $X^{k}$ are
\begin{align}
X^{+}&=x^{+}+\frac{\alpha'}{2}\,p^{+}(\sigma^{+}+\sigma^{-})
+i\sqrt{\frac{\alpha'}{2}}\sum_{n\neq 0}\frac{1}{n}
\left[
\tilde{\alpha}^{+}_{n}e^{-in\sigma^{+}}
+\alpha^{+}_{n}e^{-in\sigma^{-}}
\right]\,,
\label{solution:X+}\\
X^{k}&=x^{k}+\frac{\alpha'}{2}\,p^{k}(\sigma^{+}+\sigma^{-})
+i\sqrt{\frac{\alpha'}{2}}\sum_{n\neq 0}\frac{1}{n}
\left[
\tilde{\alpha}^{k}_{n}e^{-in\sigma^{+}}
+\alpha^{k}_{n}e^{-in\sigma^{-}}
\right]\,,
\label{solution:Xk}
\end{align}
where $\tilde{\alpha}^{+}_{n}$ and $\tilde{\alpha}^{k}_{n}$ are left-moving 
free modes, and ${\alpha}^{+}_{n}$ and ${\alpha}^{k}_{n}$ are right-moving 
free modes. 
These representations are the same as those of the usual free fields, and
$X^{+}$ and $X^{k}$ satisfy Eq.(\ref{eom:X+k}).
Of course, $X^{+}$ and $X^{k}$ can also be divided
into left-moving fields and right-moving fields, as in Eq.(\ref{sol:X+}).

Second, we present the free-mode representations of $Z$ and $\bar Z$,
which must satisfy the periodic boundary condition of closed string theory.
Because the factor $e^{-i\mu\tilde{X}^{+}(\tau,\sigma)}$ 
in $Z$ appearing in Eq.(\ref {sol:Z}) is transformed into
$e^{-2\pi i\cdot \mu \alpha' p^{+}}\cdot e^{-i\mu\tilde{X}^{+}(\tau,\sigma)}$
under the shift $\sigma\rightarrow\sigma+2\pi$,
$f(\sigma^{+})$ and $g(\sigma^{-})$ in $Z$ are twisted fields.
In other words, these fields satisfy the twisted boundary conditions,
\begin{align}
f(\sigma^{+}+2\pi)&=e^{2\pi i\cdot\mu\alpha' p^{+}}f(\sigma^{+}),\\
g(\sigma^{-}-2\pi)&=e^{2\pi i\cdot\mu\alpha' p^{+}}g(\sigma^{-}).
\end{align}
Moreover, in order for the  construction of the free-mode representation of $Z$
to be possible,
it is necessary that $Z$ satisfy the exact quantum condition.
In this paper, we use the momentum representation for $p^{+}$ and $x^{-}$,
which is a zero mode of $X^{-}$.
In this representation, $p^{+}$ is a real variable and $x^{-}$ is
the differential operator $-i\frac{\partial}{\partial p^{+}}$,
and hence the canonical commutation relation between
$p^{+}$ and $x^{-}$ is satisfied.
We thus obtain the following free-mode representation of $Z$:
\begin{align}
Z(\sigma^{+},\sigma^{-})&=e^{-i\mu\tilde{X}^{+}}
\bigl[
f(\sigma^{+})+g(\sigma^{-})
\bigr],
\label{solution:Z} \\
f(\sigma^{+})&=\sqrt{\alpha'}
\sum_{N=-\infty}^{\infty}\frac{A_{N}}{\sqrt{|N-\mu\alpha' p^{+}|}}
e^{-i(N-\mu\alpha' p^{+})\sigma^{+}},
\label{sol-f}\\
g(\sigma^{-})&=\sqrt{\alpha'}
\sum_{N=-\infty}^{\infty}\frac{B_{N}}{\sqrt{|N+\mu\alpha' p^{+}|}}
e^{-i(N+\mu\alpha' p^{+})\sigma^{-}}.
\label{sol-g}
\end{align}
Here, we assume that $\mu\alpha'p^{+}$ is not as integer.
(We present the free-mode representations in the case that
$\mu\alpha'p^{+}$ is an integer in Appendix B. 
That case includes the case $p^{+}=0$.)
The reason that the modes $A_N$ and $B_N$ in the free-mode representations
are divided by $\sqrt{|N\pm\mu\alpha' p^{+}|}$ is that
the equal-time canonical commutation relations between 
$X^{-}$ and other fields, 
for example $[X^{-}(\tau,\sigma),Z(\tau,\sigma')]=0$, must be satisfied,
and in particular, the commutators between the zero mode of $X^{-}$ and 
the modes of $Z$ must vanish. A proof of this is given in \S 4.
Similarly, taking the Hermitian conjugate of $Z$, 
we obtain the free-mode representation of $\bar{Z}$:
\begin{align}
\bar{Z}(\sigma^{+},\sigma^{-})&=e^{i\mu\tilde{X}^{+}}
\bigl[
\bar{f}(\sigma^{+})+\bar{g}(\sigma^{-})
\bigr],
\label{solution:bar-Z}\\
\bar{f}(\sigma^{+})&=\sqrt{\alpha'}
\sum_{N=-\infty}^{\infty}\frac{A_{N}^{\dagger}}
{\sqrt{|N-\mu\alpha' p^{+}|}}e^{i(N-\mu\alpha' p^{+})\sigma^{+}},\\
\bar{g}(\sigma^{-})&=\sqrt{\alpha'}
\sum_{N=-\infty}^{\infty}\frac{B_{N}^{\dagger}}
{\sqrt{|N+\mu\alpha' p^{+}|}}e^{i(N+\mu\alpha' p^{+})\sigma^{-}}\,.
\end{align}

Third, we present the free-mode representation of $X^{-}$.
To construct this representation, 
we divide $X^{-}_{\rm L}+X^{-}_{\rm R}$ in Eq.(\ref{sol:X-}) into
an almost free part, $X^{-}_{0}$, and a completely non-free part, 
$X^{-}_{1}$, which is constructed from $f$ and $g$.
Furthermore, we define $X^{-}_{2}$ as $\frac{i\mu}{2}(f\bar g-\bar f g)$
in Eq.(\ref{sol:X-}).
Here, $X_{0}^{-}$ satisfies the canonical commutation relations with
$X^{+}$ and the momentum of $X^{-}$, and it does not contain
the modes $A_{N}$, $B_{N}$ and $p^{+}$.
Moreover, $X^{-}_{1}$ commutes with $X^{+}$
and the momentum of $X^{-}$, 
and it contains the modes $A_{N}$, $B_{N}$ and $p^{+}$.
In this setting, $X^{-}$ must satisfy all the canonical commutation
relations with the string coordinates and the string momentum;
the important point is the canonical commutation relation
$[X^{-}(\tau,\sigma),Z(\tau,\sigma')]=0$,
from which we can determine the free-mode representation
of $X^{-}_1$.
Thus, using $X^{-}_{0}$,    $X^{-}_{1}$,  and $X^{-}_{2}$,
the free-mode representation of $X^{-}$ is obtained as
\begin{align}
X^{-}=X^{-}_{0}+X^{-}_{1}+X^{-}_{2}\,,
\label{solution:X-}
\end{align}
where
\begin{align}
X^{-}_{0}&=x^{-}+\frac{\alpha' p^{-}}{2}(\sigma^{+}+\sigma^{-})
+i\sqrt{\frac{\alpha'}{2}}\sum_{n\neq 0}
\frac{1}{n}
\left[\,
\tilde{\alpha}^{-}_{n}e^{-in\sigma^{+}}
+\alpha^{-}_{n}e^{-in\sigma^{-}}
\,\right]\,, 
\label{sol-X^{-}_{0}}\\
X^{-}_{1}&=\mu\alpha'
\sum_{N=-\infty}^{\infty}
\Bigl[\,
{\rm sgn}(N-\mu\alpha'p^{+}):A_{N}^{\dagger}A_{N}:
-{\rm sgn}(N+\mu\alpha'p^{+}):B_{N}^{\dagger}B_{N}:
\,\Bigr]
\tau\nonumber\\
&\;\;\;\; -\frac{i\mu\alpha'}{2}
\sum_{M\neq N}
\frac{1}{M-N}
\biggl[\,
\frac{M+N-2\mu\alpha'p^{+}}{\omega^{+}_{M}\omega^{+}_{N}}
A^{\dagger}_{M}A_{N}\,e^{i(M-N)\sigma^{+}}\nonumber\\
&\qquad\qquad\qquad\qquad\quad\;\;
-\frac{M+N+2\mu\alpha'p^{+}}{\omega^{-}_{M}\omega^{-}_{N}}
B_{M}^{\dagger}B_{N}\,e^{i(M-N)\sigma^{-}}
\,\biggr]\,, 
\label{sol-X^{-}_{1}}\\
X^{-}_{2}&=\frac{i\mu\alpha'}{2}
\sum_{M,N=-\infty}^{\infty}
\frac{1}{\omega^{+}_{M}\omega^{-}_{N}}
\Bigl[\,
A_M B_N^{\dagger}\,e^{-i(M-N-2\mu\alpha'p^{+})\tau-i(M-N)\sigma}
\nonumber\\
&\qquad\qquad\qquad\qquad\qquad\qquad\;
-A_M^{\dagger} B_N\,e^{i(M-N-2\mu\alpha'p^{+})\tau+i(M-N)\sigma}
\,\Bigr]\,.
\end{align}
Here we have $\omega^{\pm}_{N}=\sqrt{|N\mp\mu\alpha'p^{+}|}$, 
and  the notation $:\: :$ represents the normal ordered product,
whose definition is given in the final part of this section.
In the summation with $M\neq N$ in Eq.(\ref{sol-X^{-}_{1}}), 
$M$ and $N$ run from $-\infty$ to $\infty$, excluding $M=N$.
Note that the terms in this summation 
are not influenced by the normal ordered product. 
We can also write $X^{-}_{1}$ in $X^{-}$ in terms of $f$ and $g$ as
\begin{align}
X^{-}_{1}=&\frac{i\mu}{2}
\biggl[
\int d\sigma^{+}:\left(\bar{f}\partial_{+}f-\partial_{+}\bar{f}f\right):
-\int d\sigma^{-}:\left(\bar{g}\partial_{-}g
-\partial_{-}\bar{g} g\right):
\biggr]
-\mu J\sigma\,,
\label{X^{-}_{1}fg}
\end{align}
where 
\begin{align}
J=\frac{i}{4\pi}
\left[
\int_{0}^{2\pi}d\sigma^{+}
:\left(\bar{f}\partial_{+}f-\partial_{+}\bar{f}f\right):
+\int_{0}^{2\pi}d\sigma^{-}
:\left(\bar{g}\partial_{-}g-\partial_{-}\bar{g}g\right):
\right]\,.
\label{J}
\end{align}
Here, the integrals in Eq.(\ref{X^{-}_{1}fg}) are indefinite integrals,
and we choose the constants of integration to be zero.
In the free-mode representation, $J$ in Eq.(\ref{J}) is given by
\begin{align}
J=\alpha'\sum_{N=-\infty}^{\infty}
\Bigl[\,
{\rm sgn}(N-\mu\alpha'p^{+}):A^{\dagger}_{N}A_{N}:
+{\rm sgn}(N+\mu\alpha'p^{+}):B^{\dagger}_{N}B_{N}:
\,\Bigr].
\end{align}

We now explicitly present the commutation relations for all the modes, 
in order to demonstrate that they are perfectly free-modes:

$\bullet\,\,$ The nonvanishing commutation relations between
the modes of $X^{+}$ and $X^{-}$ are
\begin{align}
\left[x^{+},p^{-}\right]=\left[x^{-},p^{+}\right]=-i\,, \quad
\left[\tilde{\alpha}^{+}_{m},\tilde{\alpha}^{-}_{n}\right]=
\left[\alpha^{+}_{m},\alpha^{-}_{n}\right]=-m\delta_{m+n}\,.
\label{mode+-}
\end{align}

$\bullet\,\,$ The nonvanishing commutation relations between
the modes of $Z$ and $\bar{Z}$ are
\begin{align}
[A_{M},A^{\dagger}_{N}]&={\rm sgn}(M-\mu\alpha'p^{+})
\delta_{MN}\,,
\label{ComRelA}\\
[B_{M},B^{\dagger}_{N}]&={\rm sgn}(M+\mu\alpha'p^{+})
\delta_{MN}
\label{ComRelB}\,.
\end{align}

$\bullet\,\,$ The nonvanishing commutation relations between
the modes of $X^{k}$ are
\begin{align}
[x^{k},p^{l}]=i\delta^{kl}\,, \quad
[\tilde{\alpha}^{k}_{m},\tilde{\alpha}^{l}_{n}]=
[\alpha^{k}_{m},\alpha^{l}_{n}]=m\delta^{kl}\delta_{m+n}\,.
\end{align}
All the other commutators between the modes of 
$X^{+},\,X^{-},\,Z,\,\bar{Z}$ and $X^{k}$ vanish.
Among the vanishing commutators, those of 
$x^{-}$ given in Eq.(\ref{sol-X^{-}_{0}}) with $A_N$ and $B_N$ have
an especially important meaning.
We confirm in the next section that these commutation relations of the modes
are consistent with the canonical commutation relations of string coordinates.

Let us consider the normal ordering of the modes $A_{M}$ and $B_{M}$.
The definitions of creation and annihilation for these modes are determined 
by the signs of $\,M\pm\mu\alpha'p^{+}$ in the commutation relations
(\ref{ComRelA}) and (\ref{ComRelB}). 
For example, if $M$ is larger then $\mu\alpha'p^{+}$ ($M>\mu\alpha'p^{+}$), 
Eq.(\ref{ComRelA}) is positive ($[A_{M},A_{N}^{\dagger}]>0$), 
and thus in this case we find that $A_{M}$ are annihilation operators
and $A_{M}^{\dagger}$ are creation operators. 
Contrastingly, if $M$ is smaller than $\mu\alpha'p^{+}$ 
($M<\mu\alpha'p^{+}$), Eq.(\ref{ComRelA}) is negative
 ($[A_{M},A_{N}^{\dagger}]<0$), 
and thus in this case we find that $A_{M}^{\dagger}$ are annihilation
operators and $A_{M}$  are creation operators. 
We can determine the definitions of creation and annihilation
for $B_M$ and $B^{\dagger}_{M}$ similarly.
Therefore, the normal orderings of $A_{M}^{\dagger}A_{M}$ 
and $B_{M}^{\dagger}B_{M}$ are found to be
\begin{align}
:A^{\dagger}_{M}A_{M}:&=\left\{
\begin{array}{ll}
A^{\dagger}_{M}A_{M} & \; (M>\mu\alpha'p^{+})  \\
A_{M}A^{\dagger}_{M} & \; (M<\mu\alpha'p^{+})
\end{array}
\right.  
\label{Normal-order-A}\\
:B^{\dagger}_{M}B_{M}:&=\left\{
\begin{array}{ll}
B^{\dagger}_{M}B_{M} & \; (M>-\mu\alpha'p^{+}) \\
B_{M}B^{\dagger}_{M} & \; (M<-\mu\alpha'p^{+})
\end{array}
\right. 
\label{Normal-order-B}
\end{align}
Of course, the normal ordering of the modes
$\tilde{\alpha}^{\pm}_{n},\,\alpha^{\pm}_{n},\,
\tilde{\alpha}^{k}_{n}$ and $\alpha^{k}_{n}$
is exactly the same as that in the usual case of free fields.
The normal ordering plays an important role in the calculation of
the anomaly of the Virasoro algebra given in \S 5.

Finally, we comment on a free-field representation which is not
identical to the free-mode representation. 
Although $x^{-}$ is a free-mode, $x^{-}$ does not commute with
$f$ and $g$, which contain $p^{+}$, and therefore the commutators 
of $X^{-}_{0}$ with $f$ and $g$ do not vanish.
Therefore $X^{-}_{0}$ is not a free field. 
The derivative of $X^{-}_{0}$, however, is a free field,
because of the cancellation of $x^{-}$.
A field $X^{-}_{0}$ that is not a free field should be important
in the vertex operators, the physical states, and so on.


\vskip 1cm
\section{Proof of the free-mode representation}

In this section we prove that the free-mode representations appearing
in \S 3 satisfy the canonical commutation relations 
for all the covariant string coordinates.  
Because we need the canonical momentum to quantize the string coordinates,
we obtain the canonical momentum from the action (\ref{action-X})
using $P_{\mu}=\frac{\partial S}{\partial(\partial_{\tau}X^{\mu})}$,
$P_{Z}=\frac{\partial S}{\partial(\partial_{\tau}Z)}$
and $P_{\bar{Z}}=\frac{\partial S}{\partial(\partial_{\tau}\bar{Z})}$ :
\begin{align}
&P_{+}=-\frac{1}{2\pi\alpha'}
\Bigl[
\partial_{\tau}X^{-}+\frac{i\mu}{2}
(Z\partial_{\sigma}\bar{Z}
-\bar{Z}\partial_{\sigma} Z)+\mu^{2}\partial_{\tau}X^{+}Z\bar{Z}
\Bigr],\\
&P_{-}=-\frac{1}{2\pi\alpha'}\partial_{\tau}X^{+},
\qquad \qquad \qquad \:\:\:\:\:\:\:
P_{k}=\frac{1}{2\pi\alpha'}\partial_{\tau}X^{k}\,,\\
&P_{Z}=\frac{1}{4\pi\alpha'}
\bigl(
\partial_{\tau}\bar{Z}-i\mu\partial_{\sigma}X^{+}\bar{Z}
\bigr), 
\qquad \:\:\:
P_{\bar{{Z}}}=\frac{1}{4\pi\alpha'}
\bigl(
\partial_{\tau}Z+i\mu\partial_{\sigma}X^{+}Z
\bigr)\,.
\end{align}
Although the momentum appears complicated, it can be put into a simpler form
by using the fields $X^{+},\, \tilde{X}^{+},\,X^{-}_{0},\,f$ and $g$
which appear in the free-mode representation discussed in \S 3.
The field $P_{+}$ becomes the most simplified:
\begin{align}
&P_{+}=-\frac{1}{2\pi\alpha'}\partial_{\tau}X^{-}_{0}, 
\label{P+0}\\
&P_{-}=-\frac{1}{2\pi\alpha'}\partial_{\tau}X^{+}\,,
\qquad \qquad \qquad \:\:\:\:\:\:\:
P_{k}=\frac{1}{2\pi\alpha'}\partial_{\tau}X^{k}\,,\\
&P_{Z}=\frac{1}{4\pi\alpha'}e^{i\mu\tilde{X}^{+}}
\bigl(
\partial_{+}\bar{f}
+\partial_{-}\bar{g}
\bigr)\,,
\qquad \:\:\:
P_{\bar{Z}}=\frac{1}{4\pi\alpha'}e^{-i\mu\tilde{X}^{+}}
\bigl(
\partial_{+}f+\partial_{-}g
\bigr)\,.
\end{align} 
There are no constraints on the momentum, 
and thus we can quantize the string coordinates 
using the ordinary method of canonical quantization.
The canonical commutation relations are given by
\begin{align}
\left[X^{\mu}(\tau,\sigma),P_{\nu}(\tau,\sigma')\right]
&=i\delta^{\mu}_{\ \nu}\delta(\sigma-\sigma')\,,
\label{canonical1}\\
\left[X^{\mu}(\tau,\sigma),X^{\nu}(\tau,\sigma')\right]=0\,, \qquad
&\left[P_{\mu}(\tau,\sigma),P_{\nu}(\tau,\sigma')\right]=0\,.
\label{canonical2}
\end{align}

First, let us explain almost self-evident parts of the proof.
\vskip 3mm
\begin{enumerate}
\item Because $X^{k}$ and $P_{k}$ are constructed from usual free modes
even in our free-mode representation, it is evident that 
$X^{k}$ and $P_{k}$ satisfy the canonical commutation relations
between all string coordinates.
\vskip 2mm
\item Clearly, the canonical commutators between $Z(\tau,\sigma)$
and $Z(\tau,\sigma')$, between $Z(\tau,\sigma)$ and 
$P_{\bar Z}(\tau,\sigma')$ and between $P_{\bar Z}(\tau,\sigma)$
and $P_{\bar Z}(\tau,\sigma')$ vanish, because the modes
$A_N,\,B_N,\,p^{+},\,\tilde{\alpha}^{+}_{n}$ and $\alpha^{+}_{n}$
commute in the free-mode representation.
The same relations are satisfied in the case of 
the Hermitian conjugate of $Z$ and $P_{\bar Z}$,
namely $\bar Z$ and $P_Z$.
\vskip 2mm
\item Clearly, the canonical commutators between $X^{+}$ and $P_{-},\, Z,\,
\bar{Z},\, P_{z}$ and $P_{\bar Z}$, between $P_{-}$ and
$Z,\,\bar{Z},\, P_{z}$ and $P_{\bar Z}$,
between $X^{+}(\tau,\sigma)$ and $X^{+}(\tau,\sigma')$
and between $P_{-}(\tau,\sigma)$ and $P_{-}(\tau,\sigma')$ vanish,
because the modes
$A_N,\,B_N,\,p^{+},\,\tilde{\alpha}^{+}_{n}$ and $\alpha^{+}_{n}$
commute in the free-mode representation.
\vskip 2mm
\item $X^{+}$ and the fields constructed from only
$f,\,\bar{f},\, g$ and $\bar g$ commute,
because the modes
$x^{+},\,p^{+},\,\tilde{\alpha}_{n}$ and $\alpha_{n}$
commute with 
$p^{+},\,A_N,\,B_N,\,\bar{A}_{N}$ and $\bar{B}_{N}$
in the free-mode representation. 
Furthermore,
$X^{+}$ and $X^{-}_{0}$ commute, in analogy to usual free fields,
and hence $X^{+}$ and $X^{-}$ commute.
\vskip 2mm
\item Because $P_{+}$ contains only $\partial_{\tau}X^{-}_{0}$ in Eq.(\ref{P+0}),
it can be shown that
the canonical commutation relation between $X^{+}(\tau,\sigma)$ and
$P_{+}(\tau,\sigma')$ is satisfied by using Eq.(\ref{mode+-}). 
That between $X^{-}(\tau,\sigma)$ and
$P_{-}(\tau,\sigma')$ is also satisfied, because only $X^{-}_{0}$ in
$X^{-}$ is effective in $\partial_{\tau}X^{+}$, which $P_{-}$ contains.
Moreover, as in the case of usual free fields, 
the commutator between $P_{+}(\tau,\sigma)$ and $P_{-}(\tau,\sigma')$ 
vanishes, and that between 
$P_{+}(\tau,\sigma)$ and $P_{+}(\tau,\sigma')$ also vanishes.
\vskip 2mm
\item Because $P_{+}$ does not contain $x^{-}$, all the modes in $P_{+}$
and all the modes in $X^{-}$ commute, and thus
the canonical commutator between $P_{+}$
and $X^{-}$ vanishes.
\end{enumerate}
\vskip 3mm

We present all the important parts of the proof in next subsections.
In \S 4.1, we prove that our free-mode representation satisfies
all the important canonical commutation relations between
$Z,\,\bar{Z},\,P_{Z}$ and $P_{\bar Z}$.
In \S 4.2 we prove that our free-mode representation satisfies
all the canonical commutation relations between $X^{-}$ and
$Z,\,\bar{Z},\,P_{Z},\,P_{\bar Z}$ and $X^{-}$.
In addition, we prove that all the canonical commutation 
relations between $P_{+}$ and $Z,\,\bar{Z},\,P_{Z}$ and $P_{\bar Z}$ 
are satisfied in our free-mode representation.


\vskip 1cm
\subsection{Canonical commutation relations 
between $Z$, $\bar{Z}$, $P_Z$ and $P_{\bar{Z}}$}

In this subsection, we prove the equal-time canonical commutation 
relations involving $Z$. 
In particular, we prove the non-trivial commutation relations
$[Z(\tau,\sigma),P_{Z}(\tau,\sigma')]=i\delta(\sigma-\sigma')$, 
$[Z(\tau,\sigma),\bar{Z}(\tau,\sigma')]=0$ and 
$[P_{Z}(\tau,\sigma),P_{\bar{Z}}(\tau,\sigma')]=0$.
The commutation relation
$[\bar{Z}(\tau,\sigma),P_{\bar{Z}}(\tau,\sigma')]=i\delta(\sigma-\sigma')$ 
is obtained by taking 
the Hermitian conjugate of $[Z(\tau,\sigma),P_{Z}(\tau,\sigma')]
=i\delta(\sigma-\sigma')$, 
and therefore we need not calculate it directly.
Further, we prove the equal-time commutation relations
in arbitrary world-sheet time $\tau$. 
\\


\subsubsection{Proof of $[Z(\tau,\sigma),P_{Z}(\tau,\sigma')]
=i\delta(\sigma-\sigma')$}\label{Proof[Z,PZ]}

Let us calculate the commutator $[Z(\tau,\sigma),P_{Z}(\tau,\sigma')]$.
We have
\begin{align}
[Z(\tau,\sigma),P_{Z}(\tau,\sigma')]=\frac{1}{4\pi\alpha'}
e^{-i\mu[\tilde{X}^{+}(\sigma)-\tilde{X}^{+}(\sigma')]}
\left\{[f(\sigma^{+}),{\partial'}_{+}\bar{f}({\sigma'}^{+})]
        +[g(\sigma^{-}),{\partial'}_{-}\bar{g}({\sigma'}^{-})]\right\}\,,
\label{[Z,PZ]}
\end{align}
where ${\sigma'}^{\pm}=\tau\pm\sigma'$ and ${\partial'}_{\pm}=
\partial_{\tau}\pm\partial_{\sigma'}$. 
In \S 4.1, we use $\tilde{X}^{+}(\sigma)$ to represent
$\tilde{X}^{+}(\tau,\sigma)$.
The commutation relations $[f(\sigma^{+}),{\partial'}_{+}\bar{f}({\sigma'}^{+})]$ and 
$[g(\sigma^{-}),{\partial'}_{-}\bar{g}({\sigma'}^{-})]$ are
\begin{align}
[f(\sigma^{+}),{\partial'}_{+}\bar{f}({\sigma'}^{+})]
=i\alpha'\sum_{M,N=-\infty}^{\infty}&
\frac{N-\mu\alpha'p^{+}}{\sqrt{|M-\mu\alpha'p^{+}||N-\mu\alpha'p^{+}|}}
\nonumber\\
&\times[A_{M},A_{N}^{\dagger}]e^{-i(M-\mu\alpha'p^{+})\sigma^{+}
+i(N-\mu\alpha'p^{+}){\sigma'}^{+}}\,,
\label{f-df}\\
[g(\sigma^{-}),{\partial'}_{-}\bar{g}({\sigma'}^{-})]
=i\alpha'\sum_{M,N=-\infty}^{\infty}&
\frac{N+\mu\alpha'p^{+}}{\sqrt{|M+\mu\alpha'p^{+}||N+\mu\alpha'p^{+}|}}
\nonumber\\
&\times[B_{M},B_{N}^{\dagger}]e^{-i(M+\mu\alpha'p^{+})\sigma^{-}
+i(N+\mu\alpha'p^{+}){\sigma'}^{-}}\,.
\label{g-dg}
\end{align}
Using the commutation relation (\ref{ComRelA}), 
the commutation relation (\ref{f-df}) is reduced to
\begin{align}
[f(\sigma^{+}),{\partial'}_{+}\bar{f}({\sigma'}^{+})]=
i\alpha'\sum_{M=-\infty}^{\infty}e^{-i(M-\mu\alpha'p^{+})(\sigma-\sigma')}.\label{[f,pf]}
\end{align}
Similarly, using the commutation relation (\ref{ComRelB}), 
the other commutation relation (\ref{g-dg})
is reduced to
\begin{align}
[g(\sigma^{-}),{\partial'}_{-}\bar{g}({\sigma'}^{-})]
&=i\alpha'\sum_{M=-\infty}^{\infty}e^{i(M+\mu\alpha'p^{+})(\sigma-\sigma')}
\nonumber\\
&=i\alpha'\sum_{M=-\infty}^{\infty}e^{-i(M-\mu\alpha'p^{+})(\sigma-\sigma')}.\label{[g,pg]}
\end{align}
In the last line of this calculation, we have changed the notation $M$ to $-M$. 
Here, we note that we cannot use the normal formulas for the delta function
given in Eqs.(\ref{[f,pf]}) and (\ref{[g,pg]}) because of 
the twisted factor $e^{i\mu\alpha'p^{+}(\sigma-\sigma')}$,
which is not periodic, although the delta function is obtained through
the summations in Eqs.(\ref{[f,pf]}) and (\ref{[g,pg]}), respectively.
However, multiplying $e^{i\mu\alpha'p^{+}(\sigma-\sigma')}$ 
by $e^{-i\mu[\tilde{X}^{+}(\sigma)-\tilde{X}^{+}(\sigma')]}$
in the commutation relation (\ref{[Z,PZ]}), we can use the delta function, because 
the factor $e^{-i\mu[\tilde{X}^{+}(\sigma)-\tilde{X}^{+}(\sigma')
-\alpha'p^{+}(\sigma-\sigma')]}$ satisfies the periodic condition:
\begin{align}
[Z(\tau,\sigma),P_{Z}(\tau,\sigma')]&=\frac{i}{2\pi}
\sum_{M=-\infty}^{\infty}e^{-iM(\sigma-\sigma')}e^{-i\mu[\tilde{X}^{+}(\sigma)
-\tilde{X}^{+}(\sigma')-\alpha'p^{+}(\sigma-\sigma')]}\nonumber\\
&=i\delta(\sigma-\sigma')e^{-i\mu[\tilde{X}^{+}(\sigma)-\tilde{X}^{+}(\sigma')
-\alpha'p^{+}(\sigma-\sigma')]}.
\end{align}
Using the property of the delta function, the factor 
$e^{-i\mu[\tilde{X}^{+}(\sigma)
-\tilde{X}^{+}(\sigma')-\alpha'p^{+}(\sigma-\sigma')]}$ becomes $1$. 
Therefore, the commutation relation becomes 
$[Z(\tau,\sigma),P_{Z}(\tau,\sigma')]=i\delta(\sigma-\sigma')$.
\\


\subsubsection{Proof of $[Z(\tau,\sigma),\bar{Z}(\tau,\sigma')]=0$}\label{Proof[Z,bZ]}

Let us calculate the commutator $[Z(\tau,\sigma),\bar{Z}(\tau,\sigma')]$.
It is given by
\begin{align}
[Z(\tau,\sigma),\bar{Z}(\tau,\sigma')]=
e^{-i\mu[\tilde{X}^{+}(\sigma)-\tilde{X}^{+}(\sigma')]}
\left\{[f(\sigma^{+}),\bar{f}({\sigma'}^{+})]
+[g(\sigma^{-}),\bar{g}({\sigma'}^{-})]\right\}.
\end{align}
Using the commutation relation (\ref{ComRelA}), we can calculate 
$[f(\sigma^{+}),\bar{f}({\sigma'}^{+})]$:
\begin{align}
[f(\sigma^{+}),\bar{f}({\sigma'}^{+})]&=\alpha'\sum_{M,N=-\infty}^{\infty}
\frac{[A_{M},A_{N}^{\dagger}]}{\sqrt{|M-\mu\alpha'p^{+}||N-\mu\alpha'p^{+}|}}
e^{-i(M-\mu\alpha'p^{+})\sigma^{+}
+i(N-\mu\alpha'p^{+}){\sigma'}^{+}}\nonumber\\
&=\alpha'\sum_{M=-\infty}^{\infty}\frac{1}{M-\mu\alpha'p^{+}}
e^{-i(M-\mu\alpha'p^{+})(\sigma-\sigma')}.
\end{align} 
Similarly, using the commutation relation (\ref{ComRelB}), 
we can calculate $[g(\sigma^{-}),\bar{g}({\sigma'}^{-})]$:
\begin{align}
[g(\sigma^{-}),\bar{g}({\sigma'}^{-})]&=
\alpha'\sum_{M=-\infty}^{\infty}\frac{1}{M+\mu\alpha'p^{+}}
e^{i(M+\mu\alpha'p^{+})
(\sigma-\sigma')}\nonumber\\
&=-\alpha'\sum_{M=-\infty}^{\infty}\frac{1}{M-\mu\alpha'p^{+}}
e^{-i(M-\mu\alpha'p^{+})
(\sigma-\sigma')}\nonumber\\
&=-[f(\sigma^{+}),\bar{f}({\sigma'}^{+})]\label{ComRel:fg}.
\end{align}
In the second line of this calculation, we have changed the notation $M$ to $-M$.
Thus, because of the relation $[f(\sigma^{+}),\bar{f}({\sigma'}^{+})]
+[g(\sigma^{-}),\bar{g}({\sigma'}^{-})]=0$, 
the commutation relation $[Z(\tau,\sigma),\bar{Z}(\tau,\sigma')]$ becomes zero.\\


\subsubsection{Proof of $[P_{Z}(\tau,\sigma),P_{\bar{Z}}(\tau,\sigma')]=0$}
\label{Proof[PZ,PbZ]}

Let us calculate the commutator $[P_{Z}(\tau,\sigma),P_{\bar{Z}}(\tau,\sigma')]$.
It takes the form
\begin{align}
[P_{Z}(\tau,\sigma),P_{\bar{Z}}(\tau,\sigma')]=
&\frac{1}{(4\pi\alpha')^{2}}e^{-i\mu[\tilde{X}^{+}(\sigma)-\tilde{X}^{+}(\sigma')]}
\nonumber\\
&\times\Big\{[\partial_{+}\bar{f}(\sigma^{+}),\partial_{+}^{'}f(\sigma^{'+})]
+[\partial_{-}\bar{g}(\sigma^{-}),\partial_{-}^{'}g(\sigma^{'-})]\Big\}.
\end{align}
By a calculation similar to that in Eq.$(\ref{ComRel:fg})$, we obtain the following relation: 
\begin{align}
[\partial_{-}\bar{g}(\sigma^{-}),\partial_{-}^{'}g(\sigma^{'-})]
=-[\partial_{+}\bar{f}(\sigma^{+}),\partial_{+}^{'}f(\sigma^{'+})].\label{[pbg,pg]=-[pbf,pf]}
\end{align}
Thus, because of the relation (\ref{[pbg,pg]=-[pbf,pf]}), the commutation relation 
$[P_{Z}(\tau,\sigma),P_{\bar{Z}}(\tau,\sigma')]$ 
becomes zero.\\

Finally, from the results of \S\S \ref{Proof[Z,PZ]}, 
\ref{Proof[Z,bZ]} and 
\ref{Proof[PZ,PbZ]}, 
we find that the canonical quantization of $Z$ is consistent.


\vskip 1cm
\subsection{Canonical commutation relations of $X^{-}$ and $P_{+}$
with $Z,\,{\bar Z},\,P_{Z},\,P_{\bar Z}$ and $X^{-}$}

In this subsection, we prove the equal-time canonical commutation relations 
of $X^{-}$ and $P_{+}$ with $Z,\,\bar Z,\,P_{Z},\,P_{\bar Z},\,X^{-}$ and $P_{+}$.
This proof is the most important one in this paper, and it demonstrates the consistency 
of the free-mode representations. 
The commutation relations $[X^{-}(\tau,\sigma),Z(\tau,\sigma')]=0$, 
$[X^{-}(\tau,\sigma),P_{\bar{Z}}(\tau,\sigma')]=0$ and
$[X^{-}(\tau,\sigma),X^{-}(\tau,\sigma')]=0$, and their Hermitian conjugates, 
are non-trivial because $X^{-}$ manifestly contains $f$ and $g$. 
Therefore, we must prove these relations. 
Here, it is useful to prove the commutation relations in the case $\tau=0$ by
applying the Heisenberg formalism,
because there is reason to believe 
that $X^{-}$ takes an elegant form in this case.

At the end of this subsection, although they are trivial, 
we prove the relations 
$[P_{+}(\tau,\sigma),Z(\tau,\sigma')]=0$ and $[P_{+}(\tau,\sigma),P_{Z}(\tau,\sigma')]=0$ 
and their Hermitian conjugates. \\


\subsubsection{Preparation for the proof}

Before beginning the proof of the commutation relations involving $X^{-}$
in the free-mode representations,
we define new notation for the modes
in order to avoid complication in the coefficients,
such as the sign function.
We also present some useful formulas.

First, let us define new notation for the modes:
\begin{align}
\hat{A}_{N}&\equiv \frac{1}{\sqrt{|N-\mu\alpha'p^{+}|}}A_{N},\\
\hat{B}_{N}&\equiv \frac{1}{\sqrt{|N+\mu\alpha'p^{+}|}}B_{N}.
\end{align}
The new modes $\hat{A}_{N}$ and $\hat{B}_{N}$ interact with $x^{-}$, 
because they contain $p^{+}$. 
The commutation relations between 
these modes and $x^{-}$ are
\begin{align}
[x^{-},\hat{A}_{N}]&=-\frac{i\mu\alpha'}{2}\frac{1}{N-\mu\alpha'p^{+}}\hat{A}_{N},\\
[x^{-},\hat{B}_{N}]&=\frac{i\mu\alpha'}{2}\frac{1}{N+\mu\alpha'p^{+}}\hat{B}_{N}.
\end{align}

Next, let us construct the string coordinates in the case $\tau=0$ using
the Heisenberg formalism. We can describe the time evolution 
using the Hamiltonian of the system,
which is defined by
\begin{align}
H=&\frac{\alpha'}{2}(-2p^{+}p^{-}+p^{k}p^{k})
+\frac{1}{2}\sum_{n\neq 0}\left[-2(:\tilde{\alpha}^{+}_{n}
\tilde{\alpha}^{-}_{-n}:+:\alpha^{+}_{n}\alpha^{-}_{-n}:)
+:\tilde{\alpha}^{k}_{n}\tilde{\alpha}^{k}_{-n}:+:\alpha^{k}_{n}\alpha^{k}_{-n}:\right]
\nonumber\\
&\quad\quad+\sum_{N=-\infty}^{\infty}
\left[(N-\mu\alpha'p^{+})^{2}:\hat{A}^{\dagger}_{N}\hat{A}_{N}:
+(N+\mu\alpha'p^{+})^{2}:\hat{B}^{\dagger}_{N}\hat{B}_{N}:\right].
\end{align}
Thus, we obtain the relations
\begin{align}
Z(\tau,\sigma)&=e^{iH\tau}Z(\sigma)e^{-iH\tau},\ 
P_{\bar{Z}}(\tau,\sigma)=e^{iH\tau}P_{\bar{Z}}(\sigma)e^{-iH\tau},\nonumber\\
X^{-}(\tau,\sigma)&=e^{iH\tau}X^{-}(\sigma)e^{-iH\tau},
\end{align}
where 
we define $Z(\sigma)\equiv Z(\tau=0,\sigma)$, $P_{\bar{Z}}(\sigma)\equiv P_{\bar{Z}}(\tau=0,\sigma)$ 
and $X^{-}(\sigma)\equiv X^{-}(\tau=0,\sigma)$.
In addition, we have the following example of the commutation relations:
\begin{align}
[X^{-}(\tau,\sigma),Z(\tau,\sigma')]&=e^{iH\tau}[X^{-}(\sigma),Z(\sigma')]e^{-iH\tau}.
\end{align}
Therefore we have only to prove the relations
$[X^{-}(\sigma),Z(\sigma')]=0$, $[X^{-}(\sigma),P_{\bar{Z}}(\sigma')]=0$ and 
$[X^{-}(\sigma), X^{-}(\sigma')]=0$. 
In the case $\tau=0$, $Z(\sigma)$ and $P_{\bar{Z}}(\sigma)$ are given by
\begin{align}
Z(\sigma)&=e^{-i\mu\tilde{X}^{+}(\sigma)}
\left[f(\sigma)+g(-\sigma)\right],\\
P_{\bar{Z}}(\sigma)&=e^{-i\mu\tilde{X}^{+}(\sigma)}\left[\partial_{\sigma}f(\sigma)
-\partial_{\sigma}g(-\sigma)\right]\,,
\end{align}
where
\begin{align}
f(\sigma)&=\sqrt{\alpha'}\sum_{N=-\infty}^{\infty}\hat{A}_{N}e^{-i(N-\mu\alpha'p^{+})\sigma}\,,\\
g(-\sigma)&=\sqrt{\alpha'}\sum_{N=-\infty}^{\infty}\hat{B}_{N}
e^{i(N+\mu\alpha'p^{+})\sigma}\,.
\end{align}
Here, the minus sign on $g(-\sigma)$ is due to the origin of $\sigma^{-}$.

In the case $\tau=0$,
$X^{-}(\sigma)$ takes the elegant form
\begin{align}
X^{-}(\sigma)&=X^{-}_{0}(\sigma)-\frac{i\mu\alpha'}{2}
\left[U+V(\sigma)\right]\,,\\
U&=\sum_{N=-\infty}^{\infty}
\left[\hat{A}^{\dagger}_{N}\hat{B}_{-N}-\hat{B}^{\dagger}_{-N}\hat{A}_{N}\right]\,,\\
V(\sigma)&=\sum_{M\neq N}\frac{1}{M-N}\left[(M-\mu\alpha'p^{+})P_{MN}
+(N-\mu\alpha'p^{+})Q_{MN}\right]e^{i(M-N)\sigma}\,.
\end{align}
Here, $P_{MN}$ and $Q_{MN}$ are operators consisting of 
$\hat{A}_{N}$ and $\hat{B}_{N}$ 
and their Hermitian conjugates:
\begin{align}
P_{MN}=\left(\hat{A}^{\dagger}_{M}-\hat{B}^{\dagger}_{-M}\right)
\left(\hat{A}_{N}+\hat{B}_{-N}\right),\ \ 
Q_{MN}=\left(\hat{A}^{\dagger}_{M}+\hat{B}^{\dagger}_{-M}\right)
\left(\hat{A}_{N}-\hat{B}_{-N}\right).
\end{align}
Moreover, it is useful to define the operator 
\begin{align}
W=x^{-}-\frac{i\mu\alpha'}{2}U. 
\end{align}
The operator $W$ plays an important role in Eq.(\ref{ComRel:X-Z}). 
In the calculation of the commutation relations between $X^{-}_{0}$ and $f$
and between $X^{-}_{0}$ and $g$, 
only $x^{-}$ in $X^{-}_{0}$ survives. 
Adding $x^{-}$ to $U$, we can divide $X^{-}(\sigma)$ into 
a part with $\sigma$ dependence and the other part, 
and thus it is useful to define $W$.

We now present several useful commutation relations that we use 
in the following subsection.
First, the commutation relations between $W$ and the others are
\begin{align}
[W,p^{+}]&=-i,\label{ComRel:Wp+}\\
[W,\hat{A}_{N}]&=-\frac{i\mu\alpha'}{2}\frac{1}{N-\mu\alpha'p^{+}}
\left(\hat{A}_{N}-\hat{B}_{-N}\right),\ 
[W,\hat{B}_{N}]=-\frac{i\mu\alpha'}{2}\frac{1}{N+\mu\alpha'p^{+}}
\left(\hat{A}_{-N}-\hat{B}_{N}\right),
\label{ComRel:WAB}\\
[W,P_{MN}]&=-\frac{i\mu\alpha'}{M-\mu\alpha'p^{+}}P_{MN},
\quad\quad\quad\quad\quad\quad
[W,Q_{MN}]=-\frac{i\mu\alpha'}{N-\mu\alpha'p^{+}}Q_{MN}\label{ComRel:WPQ}.
\end{align}
Second, the commutation relations between 
$P_{MN}$ and $\hat{A}_{K}$,
$P_{MN}$ and $\hat{B}_{K}$,
$Q_{MN}$ and $\hat{A}_{K}$ 
and $Q_{MN}$ and $\hat{B}_{K}$ are
\begin{align}
[P_{MN},\hat{A}_{K}]&=-\frac{\delta_{MK}}{M-\mu\alpha'p^{+}}
\left(\hat{A}_{N}+\hat{B}_{-N}\right)\,,\:\:
&[Q_{MN},\hat{A}_{K}]=-\frac{\delta_{MK}}{M-\mu\alpha'p^{+}}
\left(\hat{A}_{N}-\hat{B}_{-N}\right)\,,
\label{ComRel:PQA}\\
[P_{MN},\hat{B}_{K}]&=\frac{\delta_{K,-M}}{K+\mu\alpha'p^{+}}
\left(\hat{A}_{N}+\hat{B}_{-N}\right)\,,\:\:
&[Q_{MN},\hat{B}_{K}]=-\frac{\delta_{K,-M}}{K+\mu\alpha'p^{+}}
\left(\hat{A}_{N}-\hat{B}_{-N}\right)\,.
\label{ComRel:PQB}
\end{align}
Finally, the commutation relations between $P_{MN}$ and $Q_{MN}$ are
\begin{align}
[P_{MN},Q_{KL}]&=0\label{ComRel:PQ},\\
[P_{MN},P_{KL}]&=2\left(\frac{\delta_{NK}}{N-\mu\alpha'p^{+}}P_{ML}
-\frac{\delta_{ML}}{M-\mu\alpha'p^{+}}P_{KN}\right)\label{ComRel:PP},\\
[Q_{MN},Q_{KL}]&=2\left(\frac{\delta_{NK}}{N-\mu\alpha'p^{+}}Q_{ML}
-\frac{\delta_{ML}}{M-\mu\alpha'p^{+}}Q_{KN}\right)\,.
\label{ComRel:QQ}
\end{align}
Below we prove the canonical commutation relations using these useful relations.\\


\subsubsection{Proof of $[X^{-}(\sigma),Z(\sigma')]=0$}

Let us calculate the commutation relation $[X^{-}(\sigma),Z(\sigma')]$.
We first note that $X^{-}_{0}(\sigma)$ and $\tilde{X}^{+}(\sigma')$
do not commute.
Therefore, $X^{-}_{0}(\sigma)$ and $e^{-i\mu\tilde{X}^{+}(\sigma')}$
do not commute.
Calculating this commutator, the following term survives:
\begin{align}
[X^{-}(\sigma),Z(\sigma')]=&e^{-i\mu\tilde{X}^{+}(\sigma')}
\Bigl\{
-i\mu[X^{-}_{0}(\sigma),\tilde{X}^{+}(\sigma')]
\left(f(\sigma')+g(-\sigma')\right)\nonumber\\
&\qquad\qquad\quad
+[W,f(\sigma')+g(-\sigma')]-\frac{i\mu\alpha'}{2}[V(\sigma),f(\sigma')+g(-\sigma')]
\Bigr\}\label{ComRel:X-Z},
\end{align}
where we use $W$ defined above. Using the fact that the modes 
$\tilde{\alpha}^{-}_{n}$ and $\alpha^{-}_{n}$ of $X^{-}$
commute with $f$ and $g$, we can construct the operator $W$. 
Here, the commutator between $X^{-}_{0}(\sigma)$ 
and $\tilde{X}^{+}(\sigma')$ is
\begin{align}
[X^{-}_{0}(\sigma),\tilde{X}^{+}(\sigma')]=-i\alpha'\sigma'
+\alpha'\sum_{n\neq 0}\frac{1}{n}e^{in(\sigma-\sigma')}.\label{ComRel:X-X+til}
\end{align}

First, using the relation (\ref{ComRel:WAB}), 
we can calculate the commutation relation $[W,f(\sigma')+g(-\sigma')]$.
Although this commutator is simply given by $f+g$ multiplied by the factor 
$\mu\alpha'\sigma'$, $x^{-}$ is very important in $W$:
\begin{align}
[W,f(\sigma')+g(-\sigma')]
=\mu\alpha'\sigma'\left[f(\sigma')+g(-\sigma')\right].\label{ComRel:Wf+g}
\end{align}
Second, using the commutation relations given in Eqs. (\ref{ComRel:PQA}) 
and (\ref{ComRel:PQB}), 
we can calculate the commutators 
$[V(\sigma),f(\sigma')]$ and $[V(\sigma),g(-\sigma')]$,
and we obtain
\begin{align}
[V(\sigma),f(\sigma')]=-\sqrt{\alpha'}&\sum_{M\neq N}\frac{1}{M-N}
\left\{
\left(\hat{A}_{N}+\hat{B}_{-N}\right)
+\frac{N-\mu\alpha'p^{+}}{M-\mu\alpha'p^{+}}
\left(\hat{A}_{N}-\hat{B}_{-N}\right)
\right\}\nonumber\\
&\times e^{i(M-N)(\sigma-\sigma')-i(N-\mu\alpha'p^{+})\sigma'}\,,\label{ComRel:Vf}
\end{align}
and
\begin{align}
[V(\sigma),g(-\sigma')]=-\sqrt{\alpha'}&\sum_{M\neq N}\frac{1}{M-N}
\left\{
\left(\hat{A}_{N}+\hat{B}_{-N}\right)
-\frac{N-\mu\alpha'p^{+}}{M-\mu\alpha'p^{+}}
\left(\hat{A}_{N}-\hat{B}_{-N}\right)
\right\}\nonumber\\
&\times e^{i(M-N)(\sigma-\sigma')-i(N-\mu\alpha'p^{+})\sigma'}.\label{ComRel:Vg}
\end{align}
Adding Eq.(\ref{ComRel:Vf}) to Eq.(\ref{ComRel:Vg}), 
the factor $\frac{N-\mu\alpha'p^{+}}{M-\mu\alpha'p^{+}}
(\hat{A}_{N}-\hat{B}_{-N})$ vanishes.
Therefore $[V(\sigma),f(\sigma')+g(-\sigma')]$ becomes 
\begin{align}
[V(\sigma),f(\sigma')+g(-\sigma')]=
-2\sqrt{\alpha'}\sum_{M\neq N}&\frac{1}{M-N}
\left(\hat{A}_{N}+\hat{B}_{-N}\right)\nonumber\\
&\times e^{i(M-N)(\sigma-\sigma')-i(N-\mu\alpha'p^{+})\sigma'}.
\end{align}
In addition, setting $M-N=n \:\,(n\neq 0)$, we obtain
\begin{align}
[V(\sigma),f(\sigma')+g(-\sigma')]&=-2\sum_{n\neq 0}\frac{1}{n}
e^{in(\sigma-\sigma')}
\cdot\sqrt{\alpha'}\sum_{N=-\infty}^{\infty}\left(\hat{A}_{N}+\hat{B}_{-N}\right)
e^{-i(N-\mu\alpha'p^{+})\sigma'}\nonumber\\
&=-2\sum_{n\neq 0}\frac{1}{n}e^{in(\sigma-\sigma')}
\left[f(\sigma')+g(-\sigma')\right]\label{ComRel:Vf+g}.
\end{align}
Finally, substituting Eqs.(\ref{ComRel:X-X+til}), (\ref{ComRel:Wf+g}) and
(\ref{ComRel:Vf+g}) into Eq.(\ref{ComRel:X-Z}), 
we find that the commutator $[X^{-}(\sigma),Z(\sigma')]$ is zero. \\


\subsubsection{Proof of $[X^{-}(\sigma),P_{\bar Z}(\sigma')]=0$}

Let us calculate the commutator $[X^{-}(\sigma),P_{\bar Z}(\sigma')]$.
We have
\begin{align}
[X^{-}(\sigma),P_{\bar{Z}}(\sigma')]=&e^{-i\mu\tilde{X}^{+}(\sigma')}
\Bigl\{
-i\mu[X^{-}_{0}(\sigma),\tilde{X}^{+}(\sigma')]
\left(\partial_{\sigma'}f(\sigma')-\partial_{\sigma}g(\sigma')\right)\nonumber\\
&+[W,\partial_{\sigma'}f(\sigma')-\partial_{\sigma'}g(-\sigma')]
-\frac{i\mu\alpha'}{2}[V(\sigma),\partial_{\sigma'}f(\sigma')
-\partial_{\sigma'}g(-\sigma')]
\Bigr\}.
\end{align}
This can be calculated in a manner similar to that used in the proof of 
$[X^{-}(\sigma),Z(\sigma')]=0$. 
First, the commutation relation of $W$ is similar to (\ref{ComRel:Wf+g}):
\begin{align}
[W,\partial_{\sigma'}f(\sigma')-\partial_{\sigma'}g(-\sigma')]=\mu\alpha'\sigma'
\left[\partial_{\sigma'}f(\sigma')-\partial_{\sigma'}g(-\sigma')\right].
\end{align}
The commutators $[V(\sigma),\partial_{\sigma'}f(\sigma')]$ 
and $[V(\sigma),\partial_{\sigma'}g(-\sigma')]$ 
are obtained by taking partial derivatives of Eqs.(\ref{ComRel:Vf}) and (\ref{ComRel:Vg}):
\begin{align}
[V(\sigma),\partial_{\sigma'}f(\sigma')]&=\partial_{\sigma'}[V(\sigma),f(\sigma')],\label{[V,pf]}\\ 
[V(\sigma),\partial_{\sigma'}g(-\sigma')]&=\partial_{\sigma'}[V(\sigma),g(-\sigma')]\label{[V,pg]}.
\end{align}
Thus, subtracting Eq.(\ref{[V,pg]}) from Eq.(\ref{[V,pf]}) and replacing $M-N$ by $n$, 
we obtain the following relation:
\begin{align}
[V(\sigma),\partial_{\sigma'}f(\sigma')-\partial_{\sigma'}g(-\sigma')]=
-2\sum_{n\neq 0}\frac{1}{n}e^{in(\sigma-\sigma')}
\left[\partial_{\sigma'}f(\sigma')-\partial_{\sigma'}g(-\sigma')\right]\,.
\end{align}
Thus, similarly to the case of $[X^{-}(\sigma),Z(\sigma')]$, 
the commutation relation $[X^{-}(\sigma),P_{\bar Z}(\sigma')]$ 
is found to be zero.
\\


\subsubsection{Proof of $[X^{-}(\sigma),X^{-}(\sigma')]=0$}

Let us calculate the commutation relation $[X^{-}(\tau,\sigma),X^{-}(\tau,\sigma')]$.
Here, it is useful to use the method of the Fourier series;
because $X^{-}(\sigma)$ is a periodic function 
under the shift $\sigma\rightarrow\sigma+2\pi$,  we can expand 
$X^{-}(\sigma)$ into a Fourier series, 
$X^{-}(\sigma)=\sum_{n}\hat{X}^{-}_{n}e^{in\sigma}$.
The Fourier coefficients are
\begin{align}
\hat{X}^{-}_{n}=\int_{0}^{2\pi} \frac{d\sigma}{2\pi}e^{-in\sigma}X^{-}(\sigma).
\end{align}
Therefore, the Fourier coefficients of 
the commutator are the following:
\begin{align}
[\hat{X}^{-}_{m},\hat{X}^{-}_{n}]
=\int_{0}^{2\pi}\frac{d\sigma}{2\pi}\int_{0}^{2\pi}\frac{d\sigma'}{2\pi}
e^{-im\sigma}e^{-in\sigma'}[X^{-}(\sigma),X^{-}(\sigma')].
\end{align}
We have only to prove $[\hat{X}^{-}_{m},\hat{X}^{-}_{n}]=0$ below. 
Calculating the commutator $[X^{-}(\sigma),X^{-}(\sigma')]$, 
the following term survives:
\begin{align}
[X^{-}(\sigma),X^{-}(\sigma')]
&=[W-\frac{i\mu\alpha'}{2}V(\sigma),W-\frac{i\mu\alpha'}{2}V(\sigma')]
\nonumber\\
&=\frac{i\mu\alpha'}{2}[W,V(\sigma)-V(\sigma')]-
\frac{\mu^{2}\alpha^{'2}}{4}[V(\sigma),V(\sigma')].\label{ComRel:X-X-}
\end{align}
However, using the relations (\ref{ComRel:Wp+}), (\ref{ComRel:WAB}) and
 (\ref{ComRel:WPQ}), 
the first term of Eq.(\ref{ComRel:X-X-}) is found to be zero:
\begin{align}
[W,V(\sigma)]=&\sum_{M\neq N}\frac{1}{M-N}
\Bigl[
i\mu\alpha'\left(P_{MN}+Q_{MN}\right)\nonumber\\
&\qquad\:\: +(M-\mu\alpha'p^{+})[W,P_{MN}]-(N-\mu\alpha'p^{+})[W,Q_{MN}]
\Bigr]
e^{i(M-N)\sigma}\nonumber\\
=&0\,.
\end{align}
Next, using the relations (\ref{ComRel:PQ}), (\ref{ComRel:PP}) and
(\ref{ComRel:QQ}), 
we can calculate the second $[V(\sigma),V(\sigma')]$ term 
in Eq.(\ref{ComRel:X-X-}): 
\begin{align}
[V(\sigma),V(\sigma')]=&2\sum_{M\neq N}\sum_{K\neq L}\frac{1}{M-N}
\frac{1}{K-L}e^{i(M-N)\sigma+i(K-L)\sigma'}\nonumber\\
&\times
\Bigl[
(M-\mu\alpha'p^{+})\delta_{NK}P_{ML}
-(K-\mu\alpha'p^{+})\delta_{ML}P_{KN}\nonumber\\
&\:\:+(M-\mu\alpha'p^{+})\delta_{NK}Q_{ML}
-(K-\mu\alpha'p^{+})\delta_{ML}Q_{KN}
\Bigr].
\end{align}
It seems that the commutator $[X^{-}(\sigma),X^{-}(\sigma')]$ is not zero.
However, we can prove that this Fourier coefficient is zero.
First, we have
\begin{align}
[\hat{X}^{-}_{m},\hat{X}^{-}_{n}]&=-\frac{\mu^{2}\alpha^{'2}}{4}
\int^{2\pi}_{0}\frac{d\sigma}{2\pi}\int^{2\pi}_{0}
\frac{d\sigma'}{2\pi}e^{-im\sigma}e^{-in\sigma'}
[V(\sigma),V(\sigma')]\nonumber\\
&=-\frac{\mu^{2}{\alpha'}^{2}}{2}
\sum_{M\neq N}\sum_{K\neq L}\frac{1}{M-N}\frac{1}{K-L}
\int^{2\pi}_{0}\frac{d\sigma}{2\pi}\int^{2\pi}_{0}
\frac{d\sigma'}{2\pi}e^{i(M-N-m)\sigma+i(K-L-n)\sigma'}
\nonumber\\
&\qquad\qquad\qquad\quad\times
\Bigl[
(M-\mu\alpha'p^{+})\delta_{NK}P_{ML}
-(K-\mu\alpha'p^{+})\delta_{ML}P_{KN}\nonumber\\
&\qquad\qquad\qquad\quad\:\:
+(M-\mu\alpha'p^{+})\delta_{NK}Q_{ML}
-(K-\mu\alpha'p^{+})\delta_{ML}Q_{KN}
\Bigr]\,.
\label{ComRel:X-mn}
\end{align}
Calculating this integral, the Kronecker delta appears in Eq.(\ref{ComRel:X-mn}).
Furthermore, carrying out the summation over $M$ and $K$, 
Eq.(\ref{ComRel:X-mn}) can be put into a simpler form, 
from which it can be seen to vanish:
\begin{align}
[\hat{X}^{-}_{n},\hat{X}^{-}_{m}]=&-\frac{\mu^{2}{\alpha'}^{2}}{2}\frac{1}{mn}
\Bigl[
\sum_{L=-\infty}^{\infty}\left(L+m+n-\mu\alpha'p^{+}\right)
\left(P_{L+m+n,L}+Q_{L+m+n,L}\right)\nonumber\\
&\quad\quad\quad\quad\:\:\:\:
-\sum_{N=-\infty}^{\infty}
\left(N+m+n-\mu\alpha'p^{+}\right)
\left(P_{N+m+n,N}+Q_{N+m+n,N}\right)\Bigr]\nonumber\\
=&0\,.
\end{align}
Therefore the Fourier coefficient of 
$[X^{-}(\sigma),X^{-}(\sigma')]$ is zero. 
This is consistent with the canonical quantization.\\


\subsubsection{Proof of $[P_{+}(\tau,\sigma),Z(\tau,\sigma')]=0$ 
and $[P_{+}(\tau,\sigma),P_{Z}(\tau,\sigma')]=0$}

It is necessary to note that the momentum $P_{+}$ contains 
$\partial_{\tau}X_{0}^{-}$, and that $Z$ and $P_{Z}$ contain $\tilde{X}^{+}$,
in order to prove the commutation relations 
$[P_{+}(\tau,\sigma),Z(\tau,\sigma')]=0$ and 
$[P_{+}(\tau,\sigma),P_{Z}(\tau,\sigma')]=0$. 
Although the modes of $\partial_{\tau}X_{0}^{-}$ 
commute with the modes of $f$ and $g$,
the modes of $\partial{\tau}X^{-}_{0}$ 
do not commute with the modes of $\tilde{X}^{+}$, 
and thus the commutation relation becomes
\begin{align}
[P_{+}(\tau,\sigma),Z(\tau,\sigma')]=-\frac{1}{2\pi\alpha'}
[\partial_{\tau}X^{-}_{0}(\tau,\sigma),\tilde{X}^{+}(\tau,\sigma')]Z(\tau,\sigma').
\end{align}
Therefore we have only to prove $[\partial_{\tau}X^{-}_{0}
(\tau,\sigma),\tilde{X}^{+}(\tau,\sigma')]=0$. 
The commutation relation becomes
\begin{align}
[\partial_{\tau}X^{-}_{0}(\tau,\sigma),\tilde{X}^{+}(\tau,\sigma')]
=i\frac{\alpha'}{2}\sum_{n\neq 0}\left[e^{in(\sigma-\sigma')}
-e^{-in(\sigma-\sigma')}\right]\,.
\end{align}
Replacing $n$ with $-n$ in the first term on the right-hand side of this equation, 
the commutator is found to be zero, and hence the commutator
$[P_{+}(\tau,\sigma),Z(\tau,\sigma')]$ is zero. 
Using the same type of calculation, we can also prove 
the relation $[P_{+}(\tau,\sigma),P_{Z}(\tau,\sigma')]=0$. 
In addition, taking the Hermitian conjugates of these commutation relations, 
we can also prove $[P_{+}(\tau,\sigma),\bar{Z}(\tau,\sigma')]=0$ and 
$[P_{+}(\tau,\sigma),P_{\bar{Z}}(\tau,\sigma')]=0$ .\\

We have thus completed the proofs of all the equal-time canonical 
commutation relations in the free-mode representations.


\vskip 1cm
\section{The energy-momentum tensor, Virasoro algebra and Virasoro anomaly}

In this section we define the energy-momentum tensors and the Virasoro operators 
using the normal procedure. 
Moreover, we exactly calculate the commutators between 
the Virasoro operators and obtain the Virasoro anomaly.

First, the variation of the action with respect to the world-sheet metric $g^{ab}$ 
defines the energy momentum tensor. 
Because the term containing $B_{\mu\nu}$ in the action (\ref{Action-X-first})
does not have the world-sheet metric, 
this term does not influence the energy-momentum tensor.
After the variation, we fix the covariant gauge as $g^{ab}=\eta^{ab}$.
The energy-momentum tensor of the string coordinates, $T^{X}_{ab}$, 
and that of the ghosts, $T^{\rm gh}_{ab}$, generally take the following forms:
\begin{align}
&T^{X}_{ab}=\frac{1}{\alpha'}G_{\mu\nu}
\Big[\partial_{a}X^{\mu}\partial_{b}X^{\nu}
-\frac{1}{2}\eta_{ab}\partial_{c}X^{\mu}\partial^{c}X^{\nu}\Big],\\
&T^{\rm gh}_{ab}=2i\Big[b_{ac}\partial_{b}c^{c}+b_{ca}\partial^{c}c_{b}
+\frac{1}{2}\partial_{c}b_{ab}c^{c}-\frac{1}{2}\eta_{ab}b_{cd}
\partial^{c}c^{d}\Big].
\end{align}
Using the world-sheet light-cone coordinates, the energy-momentum tensors
become simpler.
Moreover, substituting the solutions for the string coordinates, 
(\ref{solution:X+}), (\ref{solution:Xk}), (\ref{solution:Z}),
(\ref{solution:bar-Z}) and (\ref{solution:X-}),
and those for the ghosts, (\ref{solution:c}) and (\ref{solution:b}), 
into these energy-momentum tensors, they become
\begin{align}
&T_{++}^{X}=\frac{1}{\alpha'}\Big[-2:\partial_{+}X^{+}\partial_{+}X^{-}_{0}:
+:\partial_{+}\bar{f}\partial_{+}f:+\mu\partial_{+}X^{+}J
+:\partial_{+}X^{k}\partial_{+}X^{k}:\Big],\\
&T_{--}^{X}=\frac{1}{\alpha'}\Big[-2:\partial_{-}X^{+}\partial_{-}X^{-}_{0}:
+:\partial_{-}\bar{g}\partial_{-}g:-\mu\partial_{+}X^{+}J
+:\partial_{-}X^{k}\partial_{-}X^{k}:\Big],\\
&T_{\pm\pm}^{\rm gh}=2i\Big[:b_{\pm\pm}\partial_{\pm}c^{\pm}:
+\frac{1}{2}:\partial_{\pm}b_{\pm\pm}c^{\pm}:\Big].
\end{align}
We define the total energy-momentum tensor as $T_{\pm\pm}
\equiv T_{\pm\pm}^{X}+T_{\pm\pm}^{\rm gh}$.
Although the interaction appears in the term $\partial_{+}X^{+}J$, 
we can calculate the anomaly in almost the same manner 
as the free fields.
When we calculate the anomaly, we divide the fields into
those which consist of creation operators and 
those which consist of annihilation operators.
Here we assume $0<\mu\alpha'p^{+}<1$.

Next, we define the Virasoro operators, which are 
the Fourier coefficients of the energy-momentum tensors:
\begin{align}
\tilde{L}^{X}_{n}&=\int_{0}^{2\pi}\frac{d\sigma}{2\pi}
e^{in\sigma}T^{X}_{++},\qquad\quad
L^{X}_{n}=\int_{0}^{2\pi}\frac{d\sigma}{2\pi}e^{-in\sigma}T^{X}_{--}\,, \\
\tilde{L}^{\rm gh}_{n}&=\int_{0}^{2\pi}\frac{d\sigma}{2\pi}
e^{in\sigma}T^{\rm gh}_{++},\qquad\quad
L^{\rm gh}_{n}=\int_{0}^{2\pi}\frac{d\sigma}{2\pi}e^{-in\sigma}T^{\rm gh}_{--}\,.
\end{align}
The Virasoro operator $\tilde{L}^{X}_{n}$  ($L^{X}_{n}$) can be divided 
into $\tilde{L}_{n}^{(+-k)}$ ($L_{n}^{(+-k)}$), 
which is a part of $X^{+}$, $X^{-}_{0}$ and $X^{k}$,
and $\tilde{L}^{f}_{n}$ ($L^{g}_{n}$), which is a part of $f$ ($g$), as
$\tilde{L}_{n}^{X}=\tilde{L}_{n}^{(+-k)}+\tilde{L}_{n}^{f},\ 
L_{n}^{X}=L_{n}^{(+-k)}+L_{n}^{g}$.
In the free-mode representation, the Virasoro operators are given by
\begin{align}
\tilde{L}^{(+-k)}_{n}&=
\frac{1}{2}\sum_{m,n=-\infty}^{\infty}
\big[-2:\tilde{\alpha}^{+}_{n-m}\tilde{\alpha}^{-}_{m}:
+:\tilde{\alpha}^{k}_{n-m}\tilde{\alpha}^{k}_{m}:\big],\\         
L^{(+-k)}_{n}&=
\frac{1}{2}\sum_{m,n=-\infty}^{\infty}\big[-2:\alpha^{+}_{n-m}\alpha^{-}_{m}:
+:\alpha^{k}_{n-m}\alpha^{k}_{m}:\big], \\        
\tilde{L}^{f}_{n}&=
\frac{\mu}{\sqrt{2\alpha'}}\tilde{\alpha}^{+}_{n}J
+\sum_{N=-\infty}^{\infty}(N-n-\mu\alpha'p^{+})(N-\mu\alpha'p^{+}):
\hat{A}^{\dagger}_{N-n}\hat{A}_{N}:,\\
L^{g}_{n}&=
-\frac{\mu}{\sqrt{2\alpha'}}\alpha^{+}_{n}J
+\sum_{N=-\infty}^{\infty}(N-n+\mu\alpha'p^{+})(N+\mu\alpha'p^{+}):
\hat{B}^{\dagger}_{N-n}\hat{B}_{N}:,
\end{align}
where we define $\tilde{\alpha}^{\pm}_{0}=\alpha^{\pm}_{0}
=\sqrt{\frac{\alpha'}{2}}p^{\pm}$ and 
$\tilde{\alpha}^{k}_{0}=\alpha^{k}_{0}=\sqrt{\frac{\alpha'}{2}}p^{k}$.
Furthermore, the Virasoro operators of ghosts are given by
\begin{align}
\tilde{L}^{\rm gh}_{n}=\sum_{m=-\infty}^{\infty}(n-m):\tilde{b}_{m+n}\tilde{c}_{-m}:\,,
\qquad
L^{\rm gh}_{n}=\sum_{m=-\infty}^{\infty}(n-m):b_{m+n}c_{-m}:.
\end{align}
Calculating the commutators between the Virasoro operators 
$\tilde{L}_{n}^{f}$ and $L_{n}^{g}$, we obtain
\begin{align}
[\tilde{L}^{f}_{m},\tilde{L}^{f}_{n}]&=
(m-n)\tilde{L}^{f}_{m+n}+\tilde{A}^{f}(m)\delta_{m+n,0}\,,\\
[L^{g}_{m},L^{g}_{n}]&=
(m-n)L^{g}_{m+n}+A^{g}(m)\delta_{m+n,0}\,,
\end{align}
where $\tilde{A}^{f}(m)$ and $A^{g}(m)$ represent the anomalies of
$\tilde{L}_{n}^{f}$ and $L_{n}^{g}$.
When we calculate these anomalies, 
we must pay attention to the convergence of the infinite sum;
the anomalies are
\begin{align}
\tilde{A}^{f}(m)&=\frac{1}{6}(m^{3}-m)-\mu\alpha'p^{+}
(\mu\alpha'p^{+}-1)m,\label{anomaly:f}\\
A^{g}(m)&=\frac{1}{6}(m^{3}-m)-\mu\alpha'p^{+}(\mu\alpha'p^{+}-1)m\label{anomaly:g}.
\end{align}
Because the twisted fields $f$ and $g$ are complex fields, and each of them has 
two degrees of freedom, the coefficient $\frac{1}{6}$ appears in Eqs.(\ref{anomaly:f})
and (\ref{anomaly:g}).
In spite of the fact that the sign of the coefficient of the term 
$\mu\alpha'p^{+}$ in $\tilde{L}_{n}^{f}$ 
is different from that in $L_{n}^{g}$, 
the anomaly of $\tilde{L}_{n}^{f}$ corresponds to the anomaly of $L_{n}^{g}$. 

As a known case, the anomalies of $\tilde{L}_{n}^{(+-k)}$ 
and $L_{n}^{(+-k)}$, which contain $D-2$ string coordinates
$X^{+},\,X^{-}_{0}$ and $X^{k}$, are
$\tilde{A}^{(+-k)}(m)=A^{(+-k)}(m)=\frac{D-2}{12}(m^{3}-m)$, 
and the anomalies of $\tilde{L}_{n}^{\rm gh}$ and $L_{n}^{\rm gh}$ are
$\tilde{A}^{\rm gh}(m)=A^{\rm gh}(m)=\frac{1}{6}(m-13m^{3})$.
Let us define the total Virasoro operator in terms of an ordering constant $a$:
\begin{align}
\tilde{L}_{m}&\equiv \tilde{L}_{m}^{X}+\tilde{L}_{m}^{\rm gh}-a\delta_{m,0}\,,\\
L_{m}&\equiv L_{m}^{X}+L_{m}^{\rm gh}-a\delta_{m,0}\,.
\end{align}
Then, the Virasoro commutation relations become
\begin{align}
[\tilde{L}_{m},\tilde{L}_{n}]&=
(m-n)\tilde{L}_{m+n}+\tilde{A}(m)\delta_{m+n,0}\,,\\
[L_{m},L_{n}]&=
(m-n)L_{m+n}+A(m)\delta_{m+n,0}\,.
\end{align}
From this calculation, we find the total anomalies to be
\begin{align}
\tilde{A}(m)=A(m)=\frac{D-26}{12}(m^{3}-m)+2
\Big[a-1-\frac{1}{2}\mu\alpha'p^{+}(\mu\alpha'p^{+}-1)\Big]\,.
\label{anomaly}
\end{align}
If $D=26$ and $a=1+\frac{1}{2}\mu\alpha'p^{+}(\mu\alpha'p^{+}-1)$, 
these vanish, and the theory becomes conformally invariant.
We consider the details of this conclusion from the standpoint of BRST quantization 
in the following section.


\vskip 1cm
\section{The nilpotency of the BRST charge}
It is very important to understand the BRST quantization of string theory 
in a flat spacetime.
In this case, the nilpotency condition of the BRST charge gives
the number of spacetime dimensions $D=26$ 
and the ordering constant $a=1$.\cite{Kato-Ogawa,Hwang} 
Moreover, we have obtained 
a better understanding of BRST quantization.
For example, we now understand 
BRST cohomology, the no-ghost theorem, and the equivalence 
between BRST quantization, the old covariant quantization 
and the light-cone gauge quantization.
On the basis of these studies, it is important to understand 
the BRST quantization of string theory 
in a non-flat background. 
Here we treat the pp-wave background.
In this section, first, we define the BRST charge in terms of the Virasoro operators 
defined in the previous section. 
Second, we calculate the square of the BRST charge and impose the nilpotency 
condition on it.
Then we determine the number of dimensions of the spacetime 
and the ordering constant.

In closed string theory, the left modes and the right modes are independent.
Therefore, the BRST charge can be decomposed into left modes and right modes as
\begin{align}
Q_{\rm B}=Q^{\rm L}_{\rm B}+Q^{\rm R}_{\rm B},
\end{align}
where
\begin{align}
Q_{\rm B}^{\rm L}
&=\sum_{m=-\infty}^{\infty}:\left[\tilde{L}_{m}^{X}
+\frac{1}{2}\tilde{L}_{m}^{\rm gh}-a\delta_{m,0}\right]\tilde{c}_{-m}:\,\,,\\
Q_{\rm B}^{\rm R}
&=\sum_{m=-\infty}^{\infty}:\left[L_{m}^{X}+\frac{1}{2}L_{m}^{\rm gh}
-a\delta_{m,0}\right]c_{-m}:.
\end{align}
Concentrating our attention on the normal ordering of the mode operators, 
especially with regard to the ghosts, we obtain the square of $Q_{\rm B}$:
\begin{align}
Q_{\rm B}^{2}&=\frac{1}{2}
\Big[\left\{Q_{\rm B}^{\rm L},Q_{\rm B}^{\rm L}\right\}
+\left\{Q_{\rm B}^{\rm R},Q_{\rm B}^{\rm R}\right\}
\Big]\nonumber\\
&=\frac{1}{2}\sum_{m,n=-\infty}^{\infty}
\Big[\Big([\tilde{L}_{m},\tilde{L}_{n}]
-(m-n)\tilde{L}_{n+m}\Big)\tilde{c}_{-m}\tilde{c}_{-n}\nonumber\\
&\qquad\qquad\quad+\Big([L_{m},L_{n}]-(m-n)L_{n+m}\Big)c_{-m}c_{-n}
\Big]\nonumber\\
&=\frac{1}{2}\sum_{m=-\infty}^{\infty}A(m)\left(\tilde{c}_{-m}\tilde{c}_{m}
+c_{-m}c_{m}\right)\,,
\end{align}
where $A(m)$ is given in Eq.(\ref{anomaly}).
We note that the ghosts and the anomaly survive. 
If the anomaly is zero, the square of the BRST charge vanishes.
Because the BRST charge must have the property of nilpotency, 
the anomaly must be zero.
Thus, according to the results of the previous section, we can determine 
the number of spacetime dimensions 
and the ordering constant:
\begin{align}
D=26,\quad a=1+\frac{1}{2}\mu\alpha'p^{+}\left(\mu\alpha'p^{+}-1\right). 
\end{align}
This ordering constant corresponds to a constant that
has been determined using the method  of the $\zeta$-function 
in the light-cone gauge quantization\cite{Tseytlin95}.
However, the number of spacetime dimensions
cannot be determined in the case of the operator formalism 
of the light-cone gauge quantization\cite{F-H-H-P}.

Considering the spectrum of a closed string 
in the {\it pp}-wave background,
the physical state must satisfy $Q_{\rm B}|{\rm phys}\rangle=0$.
From this condition, we can obtain the structure of the physical state.
For example, the mass of the lightest string state, which is tachyonic,
is 
\begin{align}
m^2_{0}=-\frac{4}{\alpha'}
\left[1+\frac{1}{2}\mu\alpha'p^{+}\left(\mu\alpha'p^{+}-1\right)\right]\,,
\end{align}
and, moreover, the mass of the first excited state, which contains a massive
graviton, etc., is 
\begin{align}
m^2_{1}=-\frac{2}{\alpha'}
\mu\alpha'p^{+}\left(\mu\alpha'p^{+}-1\right)\,.
\end{align}
In the case $0<\mu\alpha'p^{+}<1$, the first excited state is stable, because
$m^2_{1}>0$; the maximum mass of this state is 
$m^2_{1}=\frac{1}{2\alpha'}$ at $\mu\alpha'p^{+}=\frac{1}{2}$.
Here, defining the mass,
we use the mass-shell condition of the particle which 
exhibits behavior similar to harmonic oscillation in non-flat directions of
the {\it pp}-wave background.


\vskip 1cm
\section{Conclusion}

In this paper we have canonically quantized a closed bosonic string in 
the {\it pp}-wave background with a non-zero $B_{\mu\nu}$ field 
using the covariant BRST operator formalism. 
In this {\it pp}-wave background,  we have constructed
the free-mode representations of all the covariant string coordinates.
Moreover, we proved that the free-mode representations
satisfy both the equal-time canonical commutation relations between
all the covariant string coordinates and the Heisenberg 
equations of motion, whose form is the same as that of the Euler-Lagrange 
equations of motion in this {\it pp}-wave background.
It is worth noting that the zero mode $x^{-}$ of $X^{-}_{0}$
has played important rules in this study.
In particular, the coefficients of the expansion modes in $Z$ and 
$\bar Z$ are determined by the condition that $x^{-}$
 must be a free mode.
 It is also interesting that $X^{-}_{0}$ is not a free field
 and the derivative of $X^{-}_{0}$ is a free field.
 Moreover, $X^{-}_{0}$ should play an important role
in the vertex operators, the physical states, and so on.

Since the energy-momentum tensor takes a very simple form 
in the free-mode representations of the covariant string coordinates,
we have been able to calculate the anomaly in the Virasoro 
algebra.
Using this anomaly, we have determined the number of dimensions of spacetime and 
the ordering constant from the nilpotency condition of the BRST charge 
in the {\it pp}-wave background.

From a new point of view regarding the free-mode representation 
in the {\it pp}-wave background, we should be able to 
obtain important information concerning
the no-ghost theorem,
all the physical states,
and the exact quantization of 
the covariant superstring.\cite{Natsuume,Berkovits}
Moreover, it would be interesting to construct 
free-mode representations in other backgrounds,
for example the shock wave.\cite{M-T-N-Y}
On the basis of these and other new concepts, 
it may be possible to elucidate the background.


 \vskip 1cm
 \section*{Acknowledgements}
 
We would like to thank H. Kunitomo for helpful comments.
This work was supported by a Grant-in-Aid for Scientific 
Research from the Japan Ministry of Education,
Culture, Sports, Science and Technology (No.13135220).


\appendix

\vskip 1cm
\section{Classical General Solution for a Closed String \\
in the pp-Wave Background without an Antisymmetric Tensor Field }

In this appendix, we consider closed string theory in the {\it pp}-wave background 
without an antisymmetric tensor field, i.e., with $B_{\mu\nu}=0$. 
In {\it pp}-wave backgrounds, the spacetime metric is generally as follows:
\begin{align}
ds^2=FdX^{+}dX^{+}-2dX^{+}dX^{-}+ dXdX+dYdY+dX^{k}dX^{k}\,\,.
\end{align}
The Einstein equation, from which the spacetime metric is obtained,
is $R_{++}=-\frac{1}{2}\left(\partial^{2}_{X}+\partial^{2}_{Y}\right)F=0$.
Here, $F$ must satisfy this Einstein equation, and therefore we choose
$F=-\mu^{2}\left(X^{2}-Y^{2}\right)$ in the case $B_{\mu\nu}$=0. 
The action is
\begin{align}
S_{X}&=-\frac{1}{4\pi\alpha'}\int d\tau d\sigma
\sqrt{-g}g^{ab}G_{\mu\nu}\partial_{a}X^{\mu}\partial_{b}X^{\nu}\,.
\end{align}
We can choose the covariant gauge $g^{ab}=\eta^{ab}$,
and then this action becomes
\begin{align}
S_{X}=-\frac{1}{4\pi\alpha'}\int d\tau d\sigma
&\Bigl[-2\partial_{a}X^{+}\partial^{a}X^{-}
-\mu^{2}\left(X^{2}-Y^{2}\right)\partial_{a}X^{+}\partial^{a}X^{+}\nonumber\\
&+\partial_{a}X\partial^{a}X
+\partial_{a}Y\partial^{a}Y
+\partial_{a}X^{k}\partial^{a}X^{k}\Bigr].
\end{align}
Therefore, the equations of motion obtained with this action are
\begin{align}
&\partial_{a}\partial^{a}X^{+}=0\,,\qquad \quad\partial_{+}\partial_{-}X^{k}=0\,,\\
&\partial_{a}\partial^{a}X=-\mu^{2}\partial_{a}X^{+}\partial^{a}X^{+}X\,,\\
&\partial_{a}\partial^{a}Y=+\mu^{2}\partial_{a}X^{+}\partial^{a}X^{+}Y\,,\\
&\partial_{a}\partial^{a}X^{-}=-\mu^{2}\partial_{a}\left[\left(X^{2}-Y^{2}\right)
\partial^{a}X^{+}\right]\,.
\end{align}
In the particle approximation, $X$ behaves similarly to a harmonic oscillator 
and $Y$ behaves similarly to an unstable operator, like a tachyon.
We can obtain classical general solutions from these equations of motion 
under the periodic boundary condition. First, the general solutions of 
$X^{+}$ and $X^{k}$ are well known:
\begin{align}
X^{+}&=x^{+}+\frac{\alpha'}{2} p^{+}(\sigma^{+}+\sigma^{-})+i\sqrt{\frac{\alpha'}{2}}
\sum_{n\neq 0}\frac{1}{n}\left[\tilde{\alpha}^{+}_{n}e^{-in\sigma^{+}}
+\alpha^{+}_{n}e^{-in\sigma^{-}}\right]\,,\\
X^{k}&=x^{k}+\frac{\alpha'}{2}\,p^{k}(\sigma^{+}+\sigma^{-})
+i\sqrt{\frac{\alpha'}{2}}\sum_{n\neq 0}\frac{1}{n}
\left[
\tilde{\alpha}^{k}_{n}e^{-in\sigma^{+}}
+\alpha^{k}_{n}e^{-in\sigma^{-}}
\right]\,.
\end{align}
Here, $X^{+}\,(X^{k})$ can be divided into  the left-moving field,
$X^{+}_{\rm L}\,(X^{k}_{\rm L})$,
and the right-moving field, $X^{+}_{\rm R}\,(X^{k}_{\rm R})$.
Second, using $X^{+}_{\rm L}$ and $X^{+}_{\rm R}$,
we can solve the equations of motion for $X$ and $Y$, 
and the general solutions under 
the periodic boundary condition of closed string theory are
\begin{align}
&X=\sqrt{\frac{\alpha'}{2}}
\sum_{N=-\infty}^{\infty}\left[{\cal A}_{N}e^{-i(\lambda_{N}X^{+}_{\rm L}
+\lambda_{-N}X^{+}_{\rm R})}
+{\cal A}^{\dagger}_{N}e^{i(\lambda_{N}X^{+}_{\rm L}
+\lambda_{-N}X^{+}_{\rm R})}\right],\\
&Y=\sqrt{\frac{\alpha'}{2}}
\sum_{N=-\infty}^{\infty}\left[{\cal B}_{N}e^{-i(\tilde{\lambda}_{N}X^{+}_{\rm L}
+\tilde{\lambda}_{-N}X^{+}_{\rm R})}
+{\cal C}_{N}e^{i(\tilde{\lambda}_{N}X^{+}_{\rm L}
+\tilde{\lambda}_{-N}X^{+}_{\rm R})}\right].
\end{align}
Here, $\lambda_{N}$ and $\tilde{\lambda}_{N}$ are defined as
\begin{align}
\lambda_{N}=\frac{1}{\alpha'p^{+}}\left[N+\sqrt{N^{2}+(\mu\alpha'p^{+})^{2}}\right]\,,
\qquad
\tilde{\lambda}_{N}=\frac{1}{\alpha'p^{+}}\left[N+\sqrt{N^{2}-(\mu\alpha'p^{+})^{2}}\right].
\end{align}
We must note that the modes ${\cal A}_{N}$ and ${\cal B}_{N}$ here are not the same as 
the modes used in the main text.
Moreover, from the relation $Y=Y^{\dagger}$, ${\cal B}_{N}$ and ${\cal C}_{N}$ must obey 
the following conditions:

$\bullet\,\,$  When $|N|>|\mu\alpha'p^{+}|,\:{\cal C}_{N}={\cal B}^{\dagger}_{N}$\,.

$\bullet\,\,$  When $|N|<|\mu\alpha'p^{+}|,\:{\cal B}^{\dagger}_{N}={\cal B}_{-N}$ and
${\cal C}^{\dagger}_{N}={\cal C}_{-N}$\,.

$\bullet\,\,$  When $|N|=|\mu\alpha'p^{+}|,\:{\cal C}_{N}={\cal B}^{\dagger}_{N}$; or 
${\cal B}^{\dagger}_{N}={\cal B}_{-N}$ and ${\cal C}^{\dagger}_{N}={\cal C}_{-N}$\,.

\noindent Next, we define 
\begin{align}
u_{N}(\tau,\sigma)
=e^{-i(\lambda_{N}X^{+}_{\rm L}+\lambda_{-N}X^{+}_{\rm R})}\,,
\qquad
v_{N}(\tau,\sigma)
=e^{-i(\tilde{\lambda}_{N}X^{+}_{\rm L}+\tilde{\lambda}_{-N}X^{+}_{\rm R})}\,.
\end{align}
Third, using world-sheet light-cone coordinates, 
the equation of motion for $X^{-}$ becomes
\begin{equation}
\partial_{+}\partial_{-}X^{-}+\frac{\mu^{2}}{2}
\left\{\partial_{+}\left[\left(X^{2}-Y^{2}\right)\partial_{-}X^{+}\right]
+\partial_{-}\left[\left(X^{2}-Y^{2}\right)\partial_{+}X^{+}\right]\right\}=0\,.
\end{equation}
Applying the inverse of the operator
 $\partial_{+}\partial_{-}$ to this equation from the left, 
we obtain the general solution of $X^{-}$,
\begin{align}
X^{-}=X^{-}_{\rm L}(\sigma^{+})+X^{-}_{\rm R}(\sigma^{-})-\frac{\mu^{2}}{2}
\left(K_{X}-K_{Y}\right)\,,
\end{align}
where $X^{-}_{\rm L}$ is a function of $\sigma^{+}$ and $X^{-}_{\rm R}$
is a function of $\sigma^{-}$. 
Moreover, $X^{-}_{\rm L}(\sigma^{+})+X^{-}_{\rm R}(\sigma^{-})$ 
and $K_{X}-K_{Y}$ are, respectively,
periodic functions of $\sigma$,
and $K_{X}$ and $K_{Y}$ are as follows:
\begin{align}
K_{X}&=\frac{\alpha'}{2}\Bigg[\sum_{N,M=-\infty}^{\infty}\frac{i}{\Lambda^{+}_{NM}}
\left({\cal A}_{N}{\cal A}_{M}u_{N}u_{M}
-{\cal A}^{\dagger}_{N}{\cal A}^{\dagger}_{M}u^{\dagger}_{N}u^{\dagger}_{M}
\right)\nonumber\\
&\qquad\:\:\:\,+\sum_{N\neq M}\frac{i}{\Lambda^{-}_{NM}}
\left({\cal A}_{N}{\cal A}^{\dagger}_{M}u_{N}u^{\dagger}_{M}
-{\cal A}^{\dagger}_{N}{\cal A}_{M}u^{\dagger}_{N}u_{M}\right)
+2\sum_{N=-\infty}^{\infty}
\left({\cal A}_{N}{\cal A}^{\dagger}_{N}
+{\cal A}^{\dagger}_{N}{\cal A}_{N}\right)X^{+}\Bigg]\,, \\
K_{Y}&=\frac{\alpha'}{2}\Biggl[\sum_{N,M=-\infty}^{\infty}\frac{i}{\tilde{\Lambda}^{+}_{NM}}
\left({\cal B}_{N}{\cal B}_{M}v_{N}v_{M}
-{\cal C}_{N}{\cal C}_{M}v^{\dagger}_{N}v^{\dagger}_{M}
\right)\nonumber\\
&\qquad\:\:\:\,+\sum_{N\neq M}\frac{i}{\tilde{\Lambda}^{-}_{NM}}
\left({\cal B}_{N}{\cal C}_{M}v_{N}v^{\dagger}_{M}
-{\cal C}_{N}{\cal B}_{M}v^{\dagger}_{N}v_{M}\right)
+2\sum_{N=-\infty}^{\infty}
\Bigl( {\cal B}_{N}{\cal C}_{N}+{\cal C}_{N}{\cal B}_{N}\Bigr) X^{+}\Biggr].
\end{align}
The quantities $\Lambda^{\pm}_{NM}$ and $\tilde{\Lambda}^{\pm}_{NM}$ here
are defined as
\begin{align}
\frac{1}{\Lambda^{\pm}_{NM}}&\equiv\frac{1}{\lambda_{N}\pm\lambda_{M}}+
\frac{1}{\lambda_{-N}\pm\lambda_{-M}},\nonumber\\ 
\frac{1}{\tilde{\Lambda}^{\pm}_{NM}}&\equiv\frac{1}{\tilde{\lambda}_{N}\pm
\tilde{\lambda}_{M}}+
\frac{1}{\tilde{\lambda}_{-N}\pm\tilde{\lambda}_{-M}}\,.
\end{align}

Finally, we consider the energy-momentum tensor in this background. 
The energy-momentum tensor is defined as 
$T_{ab}^{X}\equiv\frac{1}{\alpha'}G_{\mu\nu}
(\partial_{a}X^{\mu}\partial_{b}X^{\nu}
-\frac{1}{2}\eta_{ab}\partial_{c}X^{\mu}\partial^{c}X^{\nu})$.
Substituting the general solutions $X^{+}$, $X^{-}$, $X$, $Y$ and $X^{k}$ for
the classical energy momentum tensor $T^{X}_{ab}$, 
we obtain the following:
\begin{align}
T^{X}_{++}=\frac{1}{\alpha'}
\bigg\{&-2\partial_{+}X^{+}\partial_{+}X^{-}_{\rm L}
+\partial_{+}X^{k}\partial_{+}X^{k}\nonumber\\
&+\sum_{N=-\infty}^{\infty}\left[\Bigl(\lambda_{N}^{2}+\mu^{2}\Bigr)
\Bigl({\cal A}_{N}{\cal A}^{\dagger}_{N}+{\cal A}^{\dagger}_{N}{\cal A}_{N}\Bigr)
+\Bigl(\tilde{\lambda}_{N}^{2}-\mu^{2}\Bigr)
\Bigl({\cal B}_{N}{\cal C}_{N}+{\cal C}_{N}{\cal B}_{N}\Bigr)\right]
\big(\partial_{+}X^{+}\big)^{2}
\bigg\}\,,\\
T^{X}_{--}=\frac{1}{\alpha'}
\bigg\{&-2\partial_{-}X^{+}\partial_{-}X^{-}_{\rm R}
+\partial_{-}X^{k}\partial_{-}X^{k}\nonumber\\
&+\sum_{N=-\infty}^{\infty}\left[\Bigl(\lambda_{-N}^{2}
+\mu^{2}\Bigr)\Bigl({\cal A}_{N}{\cal A}^{\dagger}_{N}
+{\cal A}^{\dagger}_{N}{\cal A}_{N}\Bigr)
+\Bigl(\tilde{\lambda}_{-N}^{2}
-\mu^{2}\Bigr)\Bigl({\cal B}_{N}{\cal C}_{N}+{\cal C}_{N}{\cal B}_{N}\Bigr)\right]
\big(\partial_{-}X^{+}\big)^{2}
\bigg\}\,.
\end{align}
Here $T_{++}$ is a function of $\sigma^{+}$
and $T_{--}$ is a function of $\sigma^{-}$.
\\


\section{The Mode Expansion of $Z$ in the Case of $\mu\alpha'p^{+}=n$}

When $\mu\alpha'p^{+}$ is an integer ($\mu\alpha'p^{+}=n$), 
one coefficient in the mode expansion of $f$ appearing in Eq.(\ref{sol-f}) 
and one coefficient in that of $g$ appearing in Eq.(\ref{sol-g}) diverge. 
For this reason, we must consider another mode expansion.
In this case, the free-mode expansion of $Z$ becomes the following:
\begin{align}
Z(\sigma^{+},\sigma^{-})&=e^{-i\mu\tilde{X}^{+}}[f(\sigma^{+})+g(\sigma^{-})],\\
f(\sigma^{+})&=\frac{\phi_{n}}{2}+2\alpha'p_{n}\sigma^{+}+\sqrt{\alpha'}
\sum_{N\neq n}\frac{A_{N}}{\sqrt{|N-n|}}e^{-i(N-n)\sigma^{+}},\\
g(\sigma^{-})&=\frac{\phi_{n}}{2}+2\alpha'p_{n}\sigma^{-}+\sqrt{\alpha'}
\sum_{N\neq -n}\frac{B_{N}}{\sqrt{|N+n|}}e^{-i(N+n)\sigma^{-}}.
\end{align}
Of course, we can obtain the free-mode expansion of $\bar{Z}$ by taking 
the Hermitian conjugate of $Z$.
Although the above expansion slightly resembles the ordinary mode expansion 
of free fields, the above modes are complex operators, and  the coefficients
of the modes are different from those of free fields.
Moreover, $\phi_{n}$ is not the center-of-mass coordinate operator of $Z$,
and $p_{n}$ is not the total momentum operator of $Z$.
This is the case even for $p^{+}=0$.

We can canonically quantize the string coordinate $Z$, 
and when this is done
all the commutation relations between all the modes of $Z$ take
the following forms.
The commutation relations between $\phi_{n}$, $p_{n}$, $\phi_{n}^{\dagger}$
and $p_{n}^{\dagger}$ are
\begin{align}
[\phi_{n},p_{n}^{\dagger}]=i\,,\qquad [\phi_{n}^{\dagger},p_{n}]=i\,,
\end{align}
with all other commutators between them vanishing.
The commutation relations between the modes of $Z$ and $\bar{Z}$, 
except in the case
$M=\pm n$ or $N=\pm n$, correspond to Eqs.(\ref{ComRelA}) 
and (\ref{ComRelB}):
\begin{align}
[A_{M},A^{\dagger}_{N}]&={\rm sgn}(M-n)\delta_{MN},\quad (M,N\neq n)\\
[B_{M},B^{\dagger}_{N}]&={\rm sgn}(M+n)\delta_{MN},\quad (M,N\neq -n)
\end{align}
with the commutators between the other modes vanishing.
Therefore, the definitions of the normal orderings of $A_{N}$ and $B_{N}$
are obviously the same as those given 
in Eqs.(\ref{Normal-order-A}) and (\ref{Normal-order-B}).


\end{document}